\documentclass[12pt]{elsarticle}
\usepackage{refmerge}
\usepackage{epsfig}
\usepackage{lineno}
\begin{document}
\date{\today}

\title{\bf{ \boldmath
STUDY OF THE PROCESS $e^+e^-\to K_S^0 K^{\pm}\pi^{\mp} \pi^+\pi^-$
IN THE C.M. ENERGY RANGE 1.6--2.0~GEV WITH THE \mbox{CMD-3} DETECTOR
}}

\author[adr1,adr2]{R.R.~Akhmetshin}
\author[adr1,adr2]{A.N.~Amirkhanov}
\author[adr1,adr2]{A.V.~Anisenkov}
\author[adr1,adr2]{V.M.~Aulchenko}
\author[adr1]{N.S.~Bashtovoy}
\author[adr1,adr2]{D.E.~Berkaev}
\author[adr1,adr2]{A.E.~Bondar}
\author[adr1]{A.V.~Bragin}
\author[adr1,adr2,adr5]{\fbox{S.I.~Eidelman}}
\author[adr1,adr2]{D.A.~Epifanov}
\author[adr1,adr2,adr3]{L.B.~Epshteyn}
\author[adr1,adr2]{A.L.~Erofeev}
\author[adr1,adr2]{G.V.~Fedotovich}
\author[adr1,adr3]{A.O.~Gorkovenko}
\author[adr6]{F.J. Grancagnolo}
\author[adr1,adr2]{A.A.~Grebenuk}
\author[adr1,adr2]{S.S.~Gribanov}
\author[adr1,adr2,adr3]{D.N.~Grigoriev}
\author[adr1,adr2]{F.V.~Ignatov}
\author[adr1,adr2]{V.L.~Ivanov}
\author[adr1]{S.V.~Karpov}
\author[adr1,adr2]{V.F.~Kazanin}
\author[adr1]{A.N.~Kirpotin}
\author[adr1,adr2]{I.A.~Koop}
\author[adr1,adr2]{A.A.~Korobov}
\author[adr1,adr3]{A.N.~Kozyrev}
\author[adr1,adr2]{E.A.~Kozyrev}
\author[adr1,adr2]{P.P.~Krokovny}
\author[adr1,adr2]{A.S.~Kuzmin}
\author[adr1,adr2]{I.B.~Logashenko}
\author[adr1,adr2]{P.A.~Lukin}
\author[adr1,adr2]{K.Yu.~Mikhailov}
\author[adr1,adr2]{I.V.~Obraztsov}
\author[adr1]{V.S.~Okhapkin}
\author[adr1]{A.V.~Otboev}
\author[adr1]{Yu.N.~Pestov}
\author[adr1,adr2]{A.S.~Popov}
\author[adr1,adr2]{G.P.~Razuvaev}
\author[adr1,adr2]{Yu.A.~Rogovsky}
\author[adr1]{A.A.~Ruban}
\author[adr1]{N.M.~Ryskulov}
\author[adr1,adr2]{A.E.~Ryzhenenkov}
\author[adr1,adr2]{A.V.~Semenov}
\author[adr1]{A.I.~Senchenko}
\author[adr1]{P.Yu.~Shatunov}
\author[adr1]{Yu.M.~Shatunov}
\author[adr1,adr2]{V.E.~Shebalin}
\author[adr1,adr2]{D.N.~Shemyakin}
\author[adr1,adr2]{B.A.~Shwartz}
\author[adr1,adr2]{D.B.~Shwartz}
\author[adr1,adr4]{A.L.~Sibidanov}
\author[adr1,adr2]{E.P.~Solodov\fnref{tnot}}
\author[adr1]{V.M.~Titov}
\author[adr1,adr2]{A.A.~Talyshev}
\author[adr1]{S.S.~Tolmachev}
\author[adr1]{A.I.~Vorobiov}
\author[adr1]{I.M.~Zemlyansky}
\author[adr1]{D.S.~Zhadan}
\author[adr1]{A.S.~Zubakin}
\author[adr1,adr2]{Yu.V.~Yudin}

\address[adr1]{Budker Institute of Nuclear Physics, SB RAS, 
Novosibirsk, 630090, Russia}
\address[adr2]{Novosibirsk State University, Novosibirsk, 630090, Russia}
\address[adr3]{Novosibirsk State Technical University, 
Novosibirsk, 630092, Russia}
\address[adr4]{University of Victoria, Victoria, BC, Canada V8W 3P6}
\address[adr5]{Lebedev Physical Institute RAS, Moscow, 119333, Russia}
\address[adr6]{Instituto Nazionale di Fisica Nucleare, Sezione di Lecce, Lecce, Italy}
\fntext[tnot]{Corresponding author: solodov@inp.nsk.su}


%
\vspace{0.7cm}
\begin{abstract}
\hspace*{\parindent}
A cross section of the process $e^+e^- \to K_S^0 K^{\pm}\pi^{\mp}\pi^+\pi^-$ has been 
measured for the first time using a data sample of 185.4 pb$^{-1}$ collected 
with the \mbox{CMD-3} detector at the \mbox{VEPP-2000}  $e^+e^-$ collider.  With the
$K_S^0\to\pi^+\pi^-$ decay detection, 373$\pm$20 
and 514$\pm$28 signal events have been selected with six and five reconstructed 
tracks, respectively,  in the center-of-mass energy range 
1.6--2.0~GeV. The total systematic uncertainty of the cross section is
about 15\%. A study of the production dynamics allows us to extract a contribution 
from the $e^+e^- \to f_1(1285)\pi^+\pi^-$ intermediate state and to measure
the corresponding cross section. The intermediate
states with the  $f_1(1420)$ and $f_1(1510)$ resonances have been observed.
\end{abstract}

\maketitle
\baselineskip=17pt
\section{ \boldmath Introduction}
\hspace*{\parindent}
$e^+e^-$ annihilation into hadrons below 2~GeV is rich for various
multi-particle final states. Their detailed studies are important for
the development of the phenomenological models describing strong interactions
at low energies. The contributions from the different intermediate
states are particularly important for the calculations of the hadronic
vacuum polarization (HVP), and the  calculation of  the hadronic
contribution to the muon anomalous magnetic  
moment~\cite{g21,g22,g23}.
The reaction $e^+e^- \to K\bar K 3\pi$ 
has been studied before only in the $K^+K^-\pi^+\pi^-\pi^0$ mode  
by the BaBar collaboration~\cite{isr5pi},
based on the Initial-State Radiation (ISR) method. This mode is dominated by
the $\phi(1020)\eta$ and $\omega(782)K^+K^-$ intermediate states, and
possible contributions from the other states below 2~GeV are  not
observed with the available data. From the other hand, 
the $e^+e^- \to K\bar K 3\pi$ process includes many isospin
combinations of the kaons and pions, and a small signal in each mode
can give a sizable effect in the total cross section value.  
A measurement of the $e^+e^- \to K\bar K 3\pi$ cross section for the
different isospin combinations of the kaons and pions, and a study of
the  production dynamics can further improve the accuracy of the HVP 
calculations.

In this paper we report an analysis of the $e^+e^- \to K_S^0
K^{\pm}\pi^{\mp}\pi^+\pi^-$ reaction using a data sample of  
185.4 pb$^{-1}$ collected with the \mbox{CMD-3} detector 
in the E$_{\rm c.m.}$=1.6--2.0~GeV center-of-mass energy range. 
These data were collected during six energy scans, with 5--10~MeV  
c.m. energy step each, performed  at  the \mbox{VEPP-2000} $e^+e^-$ 
collider~\cite{vepp1,vepp2,vepp3,vepp4}
in 2011--2021  experimental runs. 
Starting in 2017 a beam energy was monitored by 
the back-scattering laser-light system~\cite{laser1,laser2}, providing
an absolute beam-energy measurement with better than 0.1~MeV uncertainty 
at every 10--20 minutes of the data taking. In earlier runs the beam energy 
was determined using measurements of the charged track momenta in the 
detector magnetic field with an about 1~MeV uncertainty.

The general-purpose detector \mbox{CMD-3} has been described in 
detail elsewhere~\cite{sndcmd3}. Its tracking system consists of a 
cylindrical drift chamber (DC)~\cite{dc} and double-layer multi-wire 
proportional 
Z-chamber, both are also used for a charged track trigger, and both inside 
a thin (0.2~X$_0$) superconducting solenoid with a field of 1.3~T.
The tracking system provides the (96--99)\% tracking efficiency in about 
70\% of the solid angle. The ionisation losses for the charged tracks
in the DC are measured with the 15\% accuracy.
A liquid xenon (LXe) barrel calorimeter with a 5.4~X$_0$ thickness has
a fine electrode structure, providing 1--2 mm spatial resolution for the photons 
independently of their energy~\cite{lxe}, and shares   
a cryostat vacuum volume with the solenoid.     
A barrel CsI crystal calorimeter with a thickness 
of 8.1~X$_0$ surrounds the LXe calorimeter, while an end cap BGO 
calorimeter with a thickness of 13.4~X$_0$ is placed inside the 
solenoid~\cite{cal}.
Altogether, the calorimeters cover 90\% of the solid angle and their 
amplitude signals provide information for the neutral trigger.
A charged trigger requirement of at least one track in DC is quite loose
that provides almost 100\% trigger efficiency for the process under study
with five  or six detected tracks. About 80\% of
these events have a sufficient energy deposition in the calorimeter for 
the independent neutral trigger: these events are used to control 
the charged trigger efficiency. A luminosity is measured using the 
Bhabha scattering events at large angles with about 1.5\% systematic 
uncertainty~\cite{lum}. 

To understand the detector response to the processes under study and to
obtain a detection efficiency, we have developed Monte Carlo (MC) 
simulation of our detector based on the GEANT4~\cite{geant4} package, 
in which all simulated events pass the whole reconstruction and selection 
procedure. The MC simulation uses primary generators with the matrix elements 
for the  $K_S^0 K^{\pm}\pi^{\mp}\pi^+\pi^-$ final state with the 
$f_1(1285)\rho(770)$ and $f_1(1420,1510)\rho(770)$ 
intermediate states. We simulate the $f_1(1285)$ resonance decaying to
$a_0(980)\pi$, which gives about 9\% of $f_1(1285)$ decay rate to
$K\bar K\pi$, while 
$f_1(1420,1510)$ predominantly decays to $K^*(892)K$ final state~\cite{pdg}.
The primary generator with the  $K_S^0 K^{\pm}\pi^{\mp}\pi^+\pi^-$
events uniformly distributed in the phase-space (PS)  has been also
tested. 
The primary generators include radiation of the photons by an initial 
electron and positron, calculated according to Ref.~\cite{kur_fad}. 
\section{Selection of $e^+e^-\to K_S^0 K^{\pm}\pi^{\mp}\pi^+\pi^-$ events}
\label{select}
\hspace*{\parindent}

The analysis procedure is similar to our study of the production of
six charged pions or $K_S^0 K_S^0\pi^+\pi^-$ final state described in Refs.~\cite{cmd6pi,cmd2ks2pi}. 
Candidate events
are required to have six or five tracks with the total charge zero or $\pm 1$, each having: 
\begin{itemize}
\item{
more than five hits in the DC;
}
\item{
a transverse momentum larger than 40~MeV/c;
}
\item{
a minimum distance from a track to the beam axis in the
transverse  plane of less than 6 cm, that allows reconstruction of a decay 
point of $K_S^0$ up to this distances;
}
\item{
a minimum distance from a track to the center of the interaction region along
the beam axis Z of less than 15 cm.
}
\end{itemize} 

Reconstructed momenta and angles of the tracks for the five- and six-track 
events are used for the further selection. 
  
In our reconstruction procedure we create a list of the $K_S^0\to\pi^+\pi^-$ 
candidates which includes every pair of the oppositely charged tracks, assuming 
them to be pions, with an invariant mass within $\pm 80$~MeV/c$^2$ from 
the $K_S^0$ mass~\cite{pdg}
and a common vertex point within a spacial uncertainty of the DC.  
We calculate momentum and energy for the $K_S^0$ candidate taking 
the values of the pions mass from Ref.~\cite{pdg}.

\begin{figure*}[p]
\begin{center}
\vspace{-0.5cm}
\includegraphics[width=0.5\textwidth]{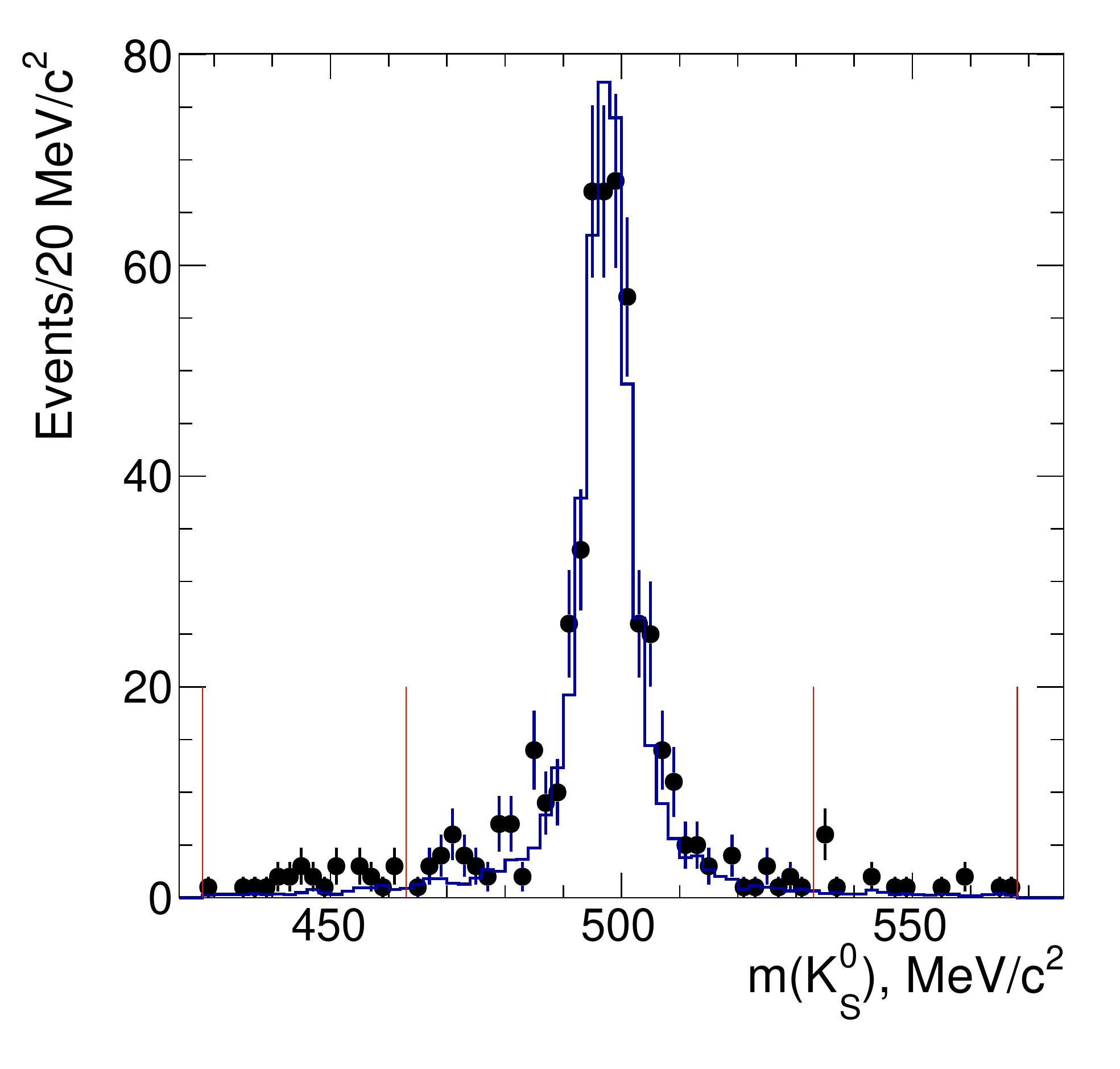}
\put(-50,150){\makebox(0,0)[lb]{\bf(a)}}
\includegraphics[width=0.49\textwidth]{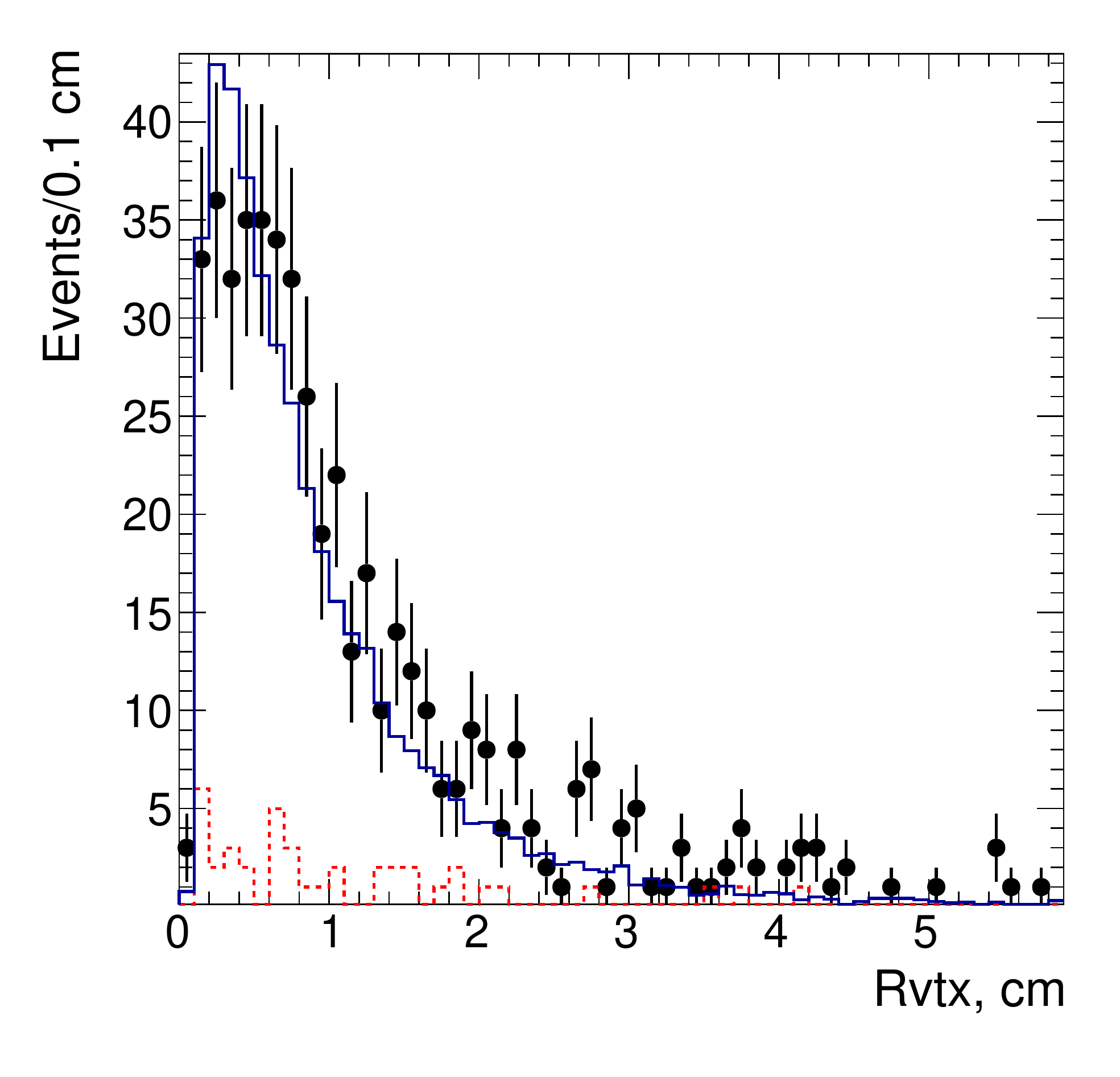}
\put(-50,150){\makebox(0,0)[lb]{\bf(b)}}
\\
\vspace{-0.4cm}
\includegraphics[width=0.5\textwidth,height=0.45\textwidth]{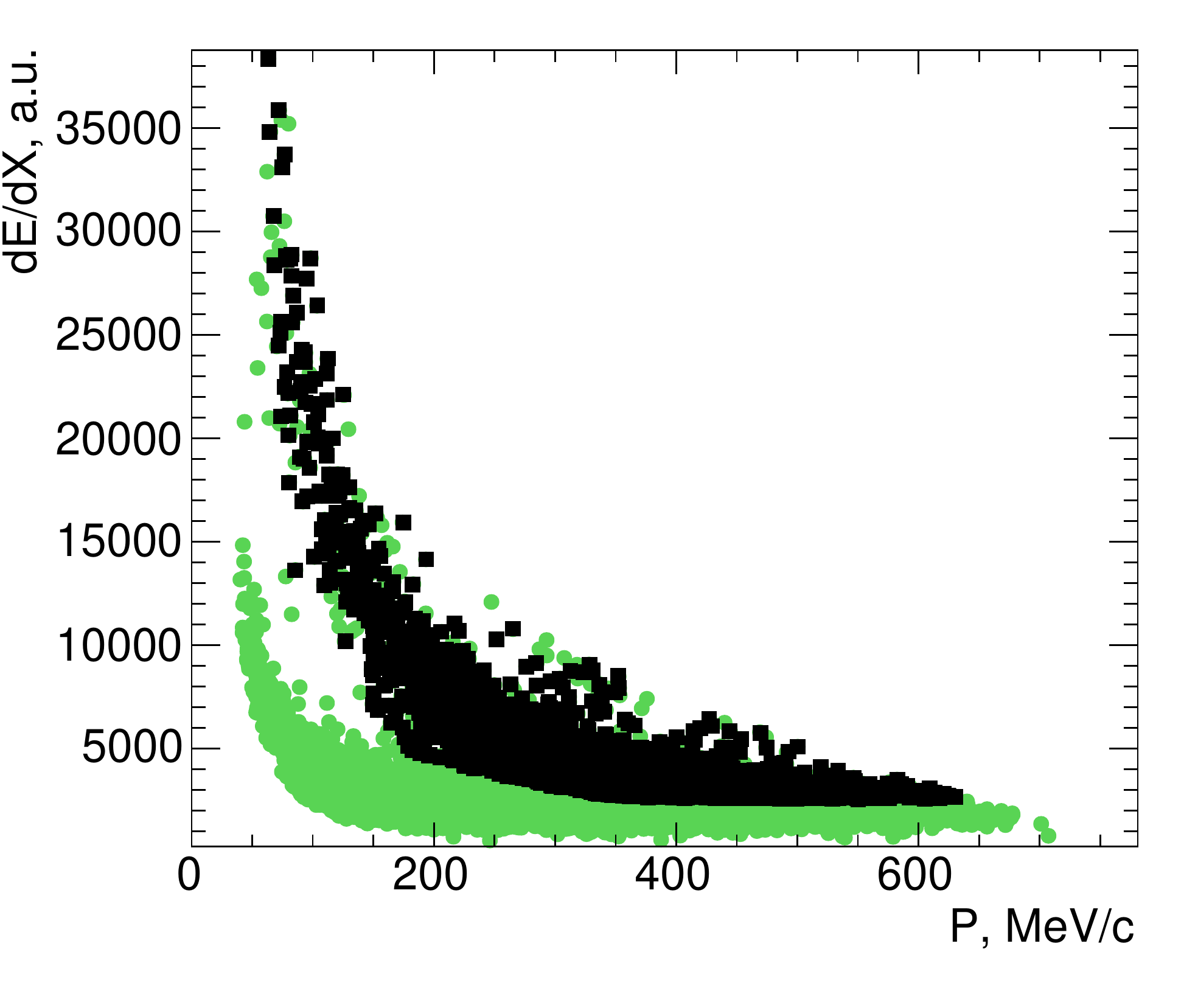}
\put(-50,140){\makebox(0,0)[lb]{\bf(c)}}
\includegraphics[width=0.49\textwidth]{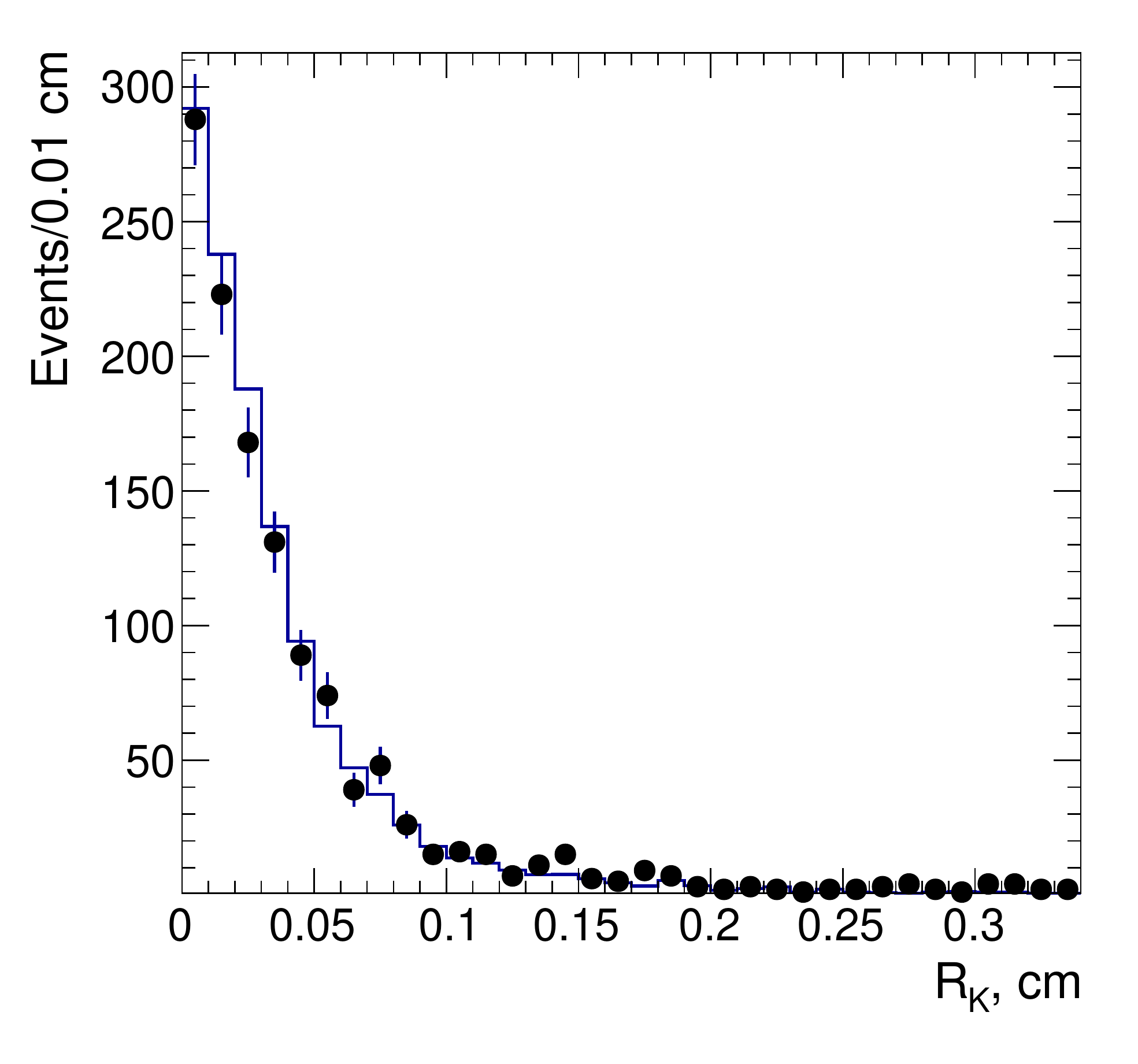}
\put(-50,140){\makebox(0,0)[lb]{\bf(d)}}
\vspace{-0.4cm}
\caption
{
  (a) The invariant mass of the $K_S^0 $ candidate for the data (dots) and simulation (histogram). The lines show
  selections for the signal events and for the control region to estimate the background level.  
  (b) The transverse distance for the $K_S^0$ candidate 
 vertices from the beam axis. Solid histogram is for simulation, dashed histogram is for
 the background.  (c) The dE/dX distribution vs momentum for all events with
 six tracks. Black squares indicate selected charged kaon candidates.
 (d) The transverse distance for the charged kaon candidate
 tracks from the beam axis for the data (dots) and simulation (histogram).
}
\label{mksk}
\end{center}
\end{figure*}

At the first stage of the signal event selection we require at least one $K_S^0$ 
candidate with two independent tracks plus three  or four additional charged 
tracks, which are required to be within 0.35 cm from the beam axis.
If there are more than one $K_S^0$ candidate, a candidate pair with a 
minimal deviation from the $K_S^0$ mass is retained. 
A charged kaon candidate is selected from these
beam-originated tracks using the ionization loss measurement in the DC as described below.

The signal event candidate is required to
have at least  one $K_S^0$ and at least one $K^{\pm}$ for the next selection stage. 

Figure~\ref{mksk}(a) shows the invariant mass of the 
$K_S^0\to\pi^+\pi^-$ candidates for the data (points) and MC  
simulation (histogram). The data from the energy intervals above
E$_{\rm c.m.}>$1950 MeV are combined for the presented histograms.
The vertical lines show selections for the signal events and events 
for a background level estimate by using the side bands with equal 
to the signal mass intervals.
The radial distances of the $K_S^0$ decay vertices from the beam axis are shown in
Fig.~\ref{mksk}(b) for the data (points) and MC simulation 
(histogram) for the events in the signal region. The dashed histogram
represents a background contribution,
estimated from the side bands of Fig.~\ref{mksk}(a).
Figure~\ref{mksk}(c) shows a scatter plot for the dE/dX values vs
momentum for the six tracks of the events from  Fig.~\ref{mksk}(a).
Clear bands for the pions and kaons are seen. 
The signals from the kaons and pions are overlapped at large momenta, and
we use  a selection boundary formed by the two linear functions.
The most critical is the horizontal boundary above the 300 MeV/c momentum.
This boundary is set at the two sigma
level above the average dE/dX value for the pions, and rejects about
50\% of the kaons in this momentum range.
The black dots show the selected  charged kaon candidates.
The radial distance from the beam axis for the charged kaon tracks is shown in Fig.~\ref{mksk}(d).
Our simulation well describes the experimental distributions.

\begin{figure}[tbh]
\begin{center}
\vspace{-0.cm}
\includegraphics[width=1.0\textwidth]{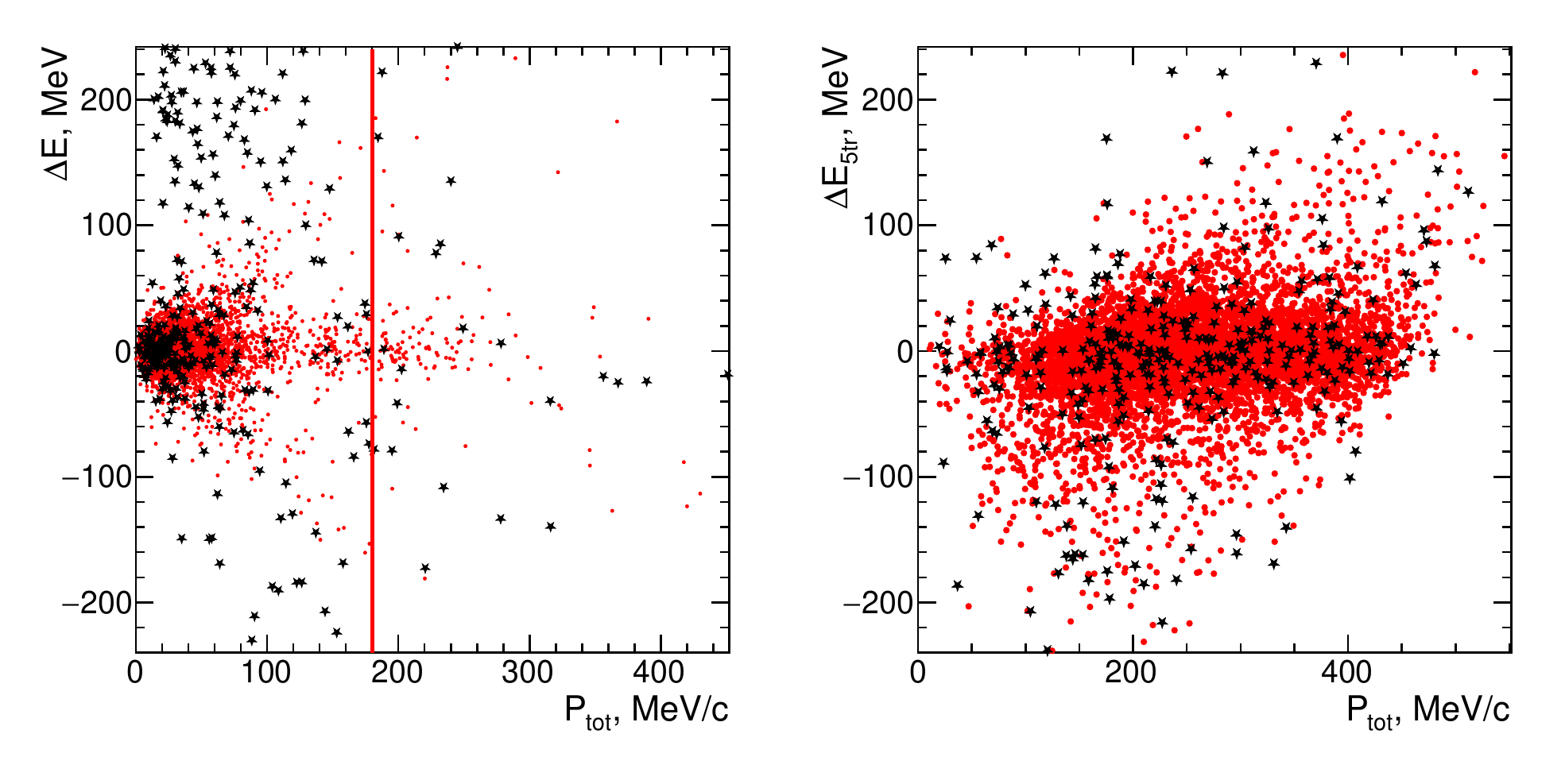}
\put(-245,153){\makebox(0,0)[lb]{\bf(a)}}
\put(-145,153){\makebox(0,0)[lb]{\bf(b)}}
\vspace{-0.7cm}
\caption
{
The scatter plot of the difference  between the energy of 
$K_S^0 K^{\pm}\pi^{\mp}\pi^+\pi^-$ candidates 
and c.m. energy vs total momentum, $\Delta \rm E$, for events with six tracks (a) and events with five tracks (b). The 
stars are for data, while the signal simulation is shown by red 
(in color version) dots; the line shows the applied selection.
}
\label{energy}
\end{center}
\end{figure}

For the six- and five-track $K_S^0 K^{\pm}\pi^{\mp}\pi^+\pi^-$ candidates we 
calculate the total energy of two kaons and three pions: for the five-track 
candidates a missing  momentum is used to calculate an energy of the lost 
pion. Figure~\ref{energy} shows a scatter plot of the difference between 
the total energy and c.m. energy, $\rm \Delta E = E_{tot}-E_{\rm c.m.}$, vs the total
momentum,  $\rm P_{\rm tot}$, of six- (a) and five-track (b) candidates.
A clear signal of the $e^+ e^- \to K_S^0 K^{\pm}\pi^{\mp}\pi^+\pi^-$ reaction is seen in 
Fig.~\ref{energy}(a) as a cluster of the stars near zero, in agreement with  
the expectation from the simulation shown by (red in the color version) dots.
We require $\rm P_{\rm tot}$ to be less than 180~MeV/c, thus reducing a number of 
signal events with the hard radiative photons.

The expected  signal of the five-track candidates has the $\rm \Delta E_{5tr}$ 
value near zero, and the $\rm P_{tot}$ value is distributed up to
about 500~MeV/c, 
as shown by the (red) dots from the signal MC simulation in 
Fig.~\ref{energy}(b). The  
(black) stars show our data: signal events are clearly seen.

\begin{figure}[tbh]
\begin{center}
\vspace{-0.cm}
\includegraphics[width=1.0\textwidth]{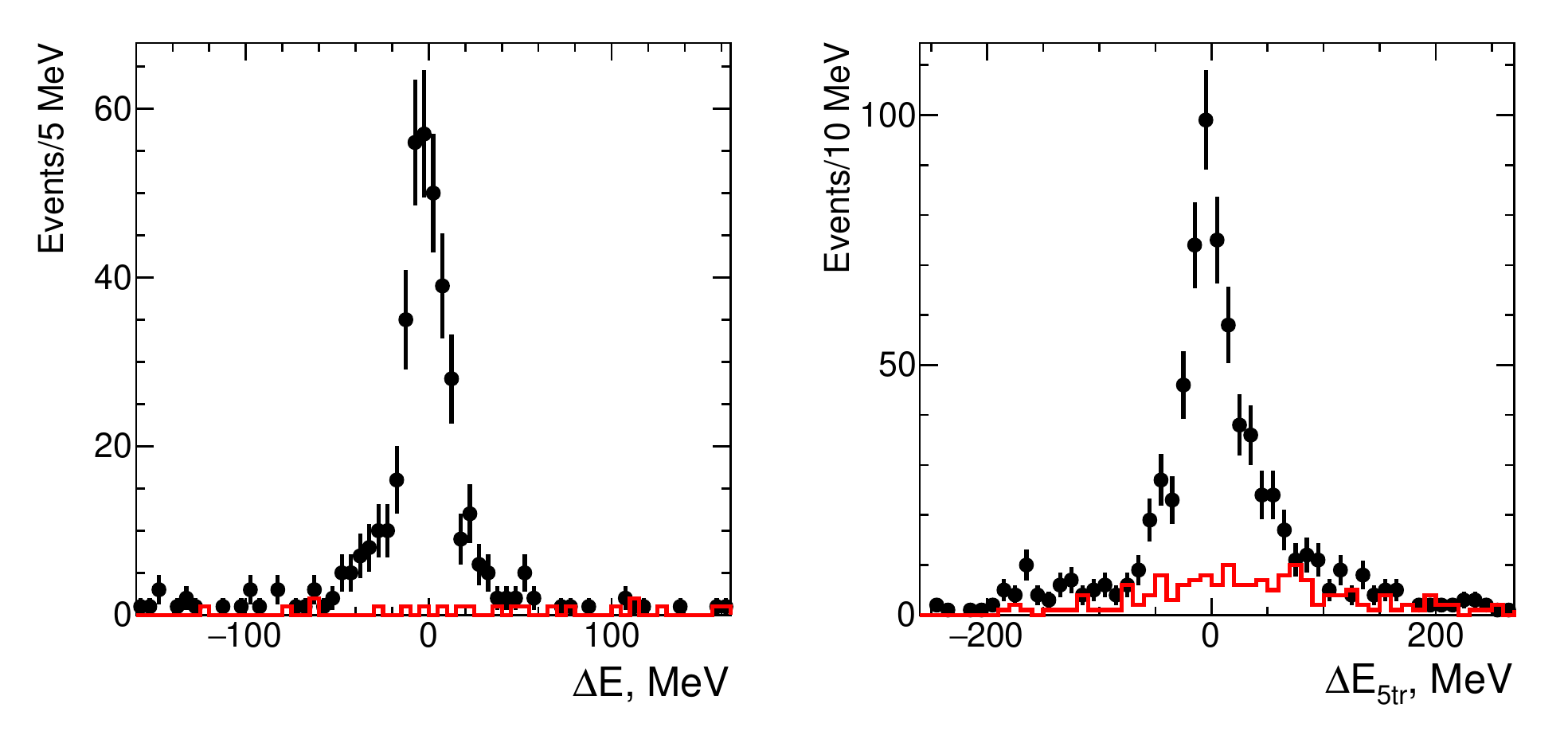}
\put(-245,153){\makebox(0,0)[lb]{\bf(a)}}
\put(-45,153){\makebox(0,0)[lb]{\bf(b)}}
\vspace{-0.8cm}
\caption
{
(a) The difference  between the energy of the $K_S^0 K^{\pm}\pi^{\mp}\pi^+\pi^-$ 
candidates and c.m. energy, $\Delta \rm E$, after selection shown by the line in Fig.~\ref{energy}(a) 
for the six-track (a)  and the five-track (b) events. All the energy intervals 
are summed. The histograms show the background contribution.
}
\label{energy1D}
\end{center}
\end{figure}

The background contribution from the $e^+ e^- \to3(\pi^+\pi^-)$ reaction,
seen as a cluster at $\Delta \rm E =$~200~MeV in Fig.~\ref{energy}(a), is 
effectively reduced by a requirement of the minimum distance between any 
track from the $K_S^0$ decay and the beam axis to be larger than 0.1 cm. 
A signal loss due to this requirement is small as seen in Fig.~\ref{mksk}(b).
Another significant background from the $e^+ e^- \to K_S^0 K^{\pm}\pi^{\mp}\pi^0$
reaction with additional tracks from the photon conversion  is 
reduced by  a requirement of missing mass to any of the
$K_S^0 K^{\pm}\pi^{\mp}$ combination to be larger than two pion masses.

Figure~\ref{energy1D} shows the projection plots of Fig.~\ref{energy}, 
$\rm \Delta E$, for the six-track (a)  
and the five-track (b) events with all above selection applied: the dots present events from 
the signal region, while the histograms are the events from the side bands
of Fig.~\ref{mksk}(a), which are used to subtract the remaining  background 
contribution. All energy intervals are summed. The background contribution is small: almost negligible
for the six-track events and is about 15\% for the five-track events.

As seen in Fig.~\ref{mksk}(c), the dE/dX values for the kaons and pions
are significantly overlapped at large momentum, 
and above the selection boundary we observe about 20\% of the events with more than one charged kaon candidate.
In this case we apply kaon mass to only one candidate, assume all the other 
tracks to be pions, test all combinations, and retain a combination with 
the $\rm\Delta E$  closest to zero. The MC simulation shows that the procedure
relatively well recover the leakage of the pions to the kaon selections.

\begin{figure}[p]
\begin{center}
\vspace{0.1cm}
\includegraphics[width=0.9\textwidth]{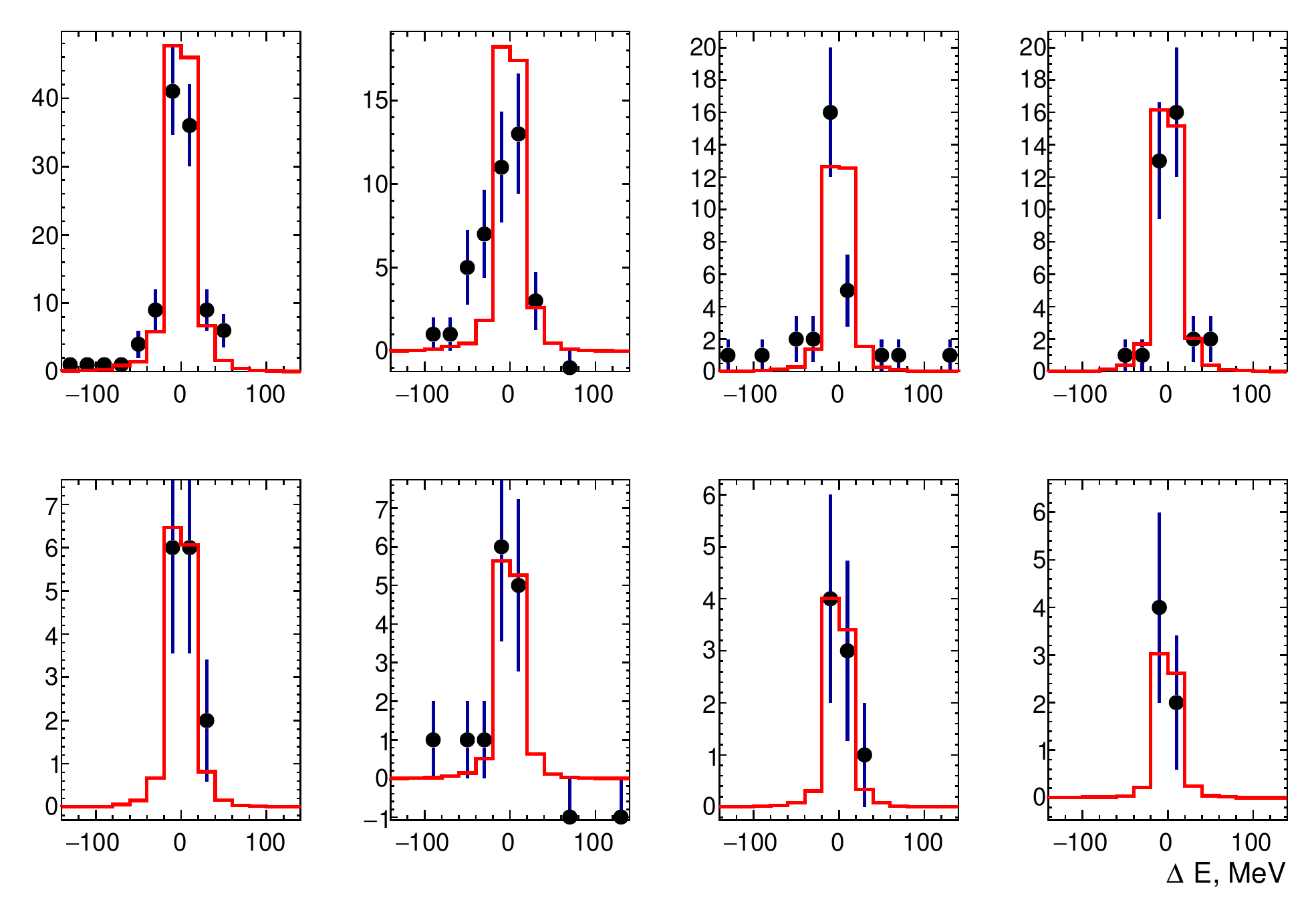}
\put(-285,235){\makebox(0,0)[lb]{\bf(a)}}
\hfill
\includegraphics[width=0.9\textwidth]{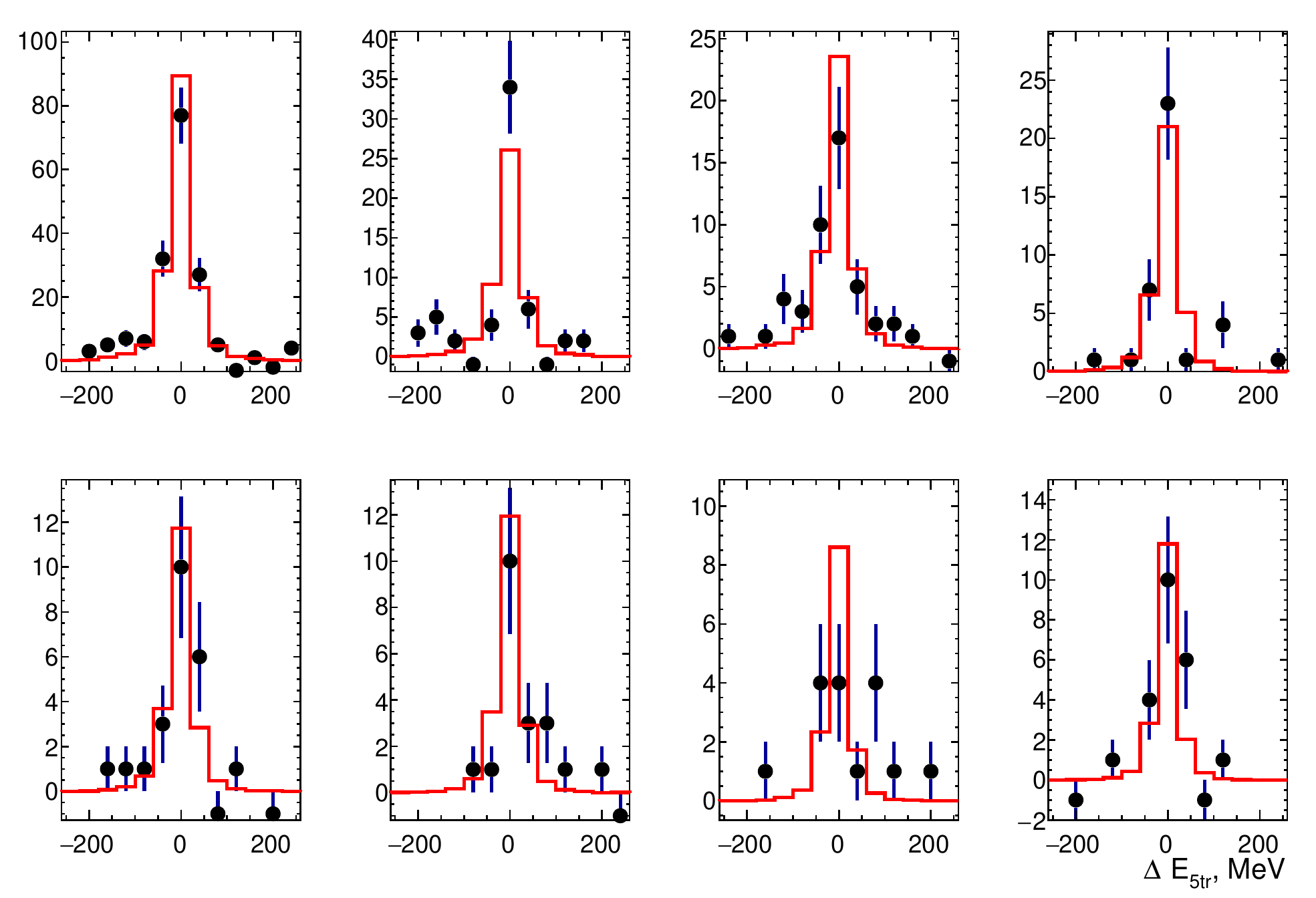}
\put(-285,235){\makebox(0,0)[lb]{\bf(b)}}
\vspace{-0.3cm}
\caption
{
(a) The difference  between the energy of the $K_S^0 K^{\pm}\pi^{\mp}\pi^+\pi^-$ 
candidates and c.m. energy, $\Delta \rm E$, after background subtraction for six-track  (a) and five-track 
events (b) for eight c.m. energy intervals (dots): left to right, 
top to bottom according to the last eight lines of Table~\ref{xs_all}. 
Histograms show expected signals from simulation, normalized to the 
total number of events in each plot.
}
\label{6energy}
\end{center}
\end{figure}

The observed signal is small, and to get a reasonable number of events we combine our scanned 
points from the early runs into eight energy intervals, as shown in
Table~\ref{xs_all}, while the  latest scans with the larger integrated
luminosity are presented as individual energy points.  
To obtain the number of signal events, we use $\rm \Delta E$ distributions of 
Fig.~\ref{energy1D} after the background subtraction
for each energy interval for the six- and the five-track events. We  count remaining 
events in the $\pm$70~MeV region for the six-track events, and in the 
$\pm$150~MeV region for the five-track events. The obtained differences are 
shown in Fig.~\ref{6energy} by dots for the six- (a) and the five-track (b) events: 
from left to right, from top to bottom according to the last eight energy points of 
Table~\ref{xs_all}. The histograms show expected signals from the simulation. 
In total, we obtain 373$\pm$20 and 514$\pm$28  six- and 
five-track signal events, respectively. 
The numbers of selected events in each energy interval are 
listed in Table~\ref{xs_all}.

\section{Study of the production dynamics}
\label{dynamics}
\hspace*{\parindent}

\begin{figure}[tbh]
\vspace{-0.cm}
\includegraphics[width=0.33\textwidth,height=0.33\textwidth]{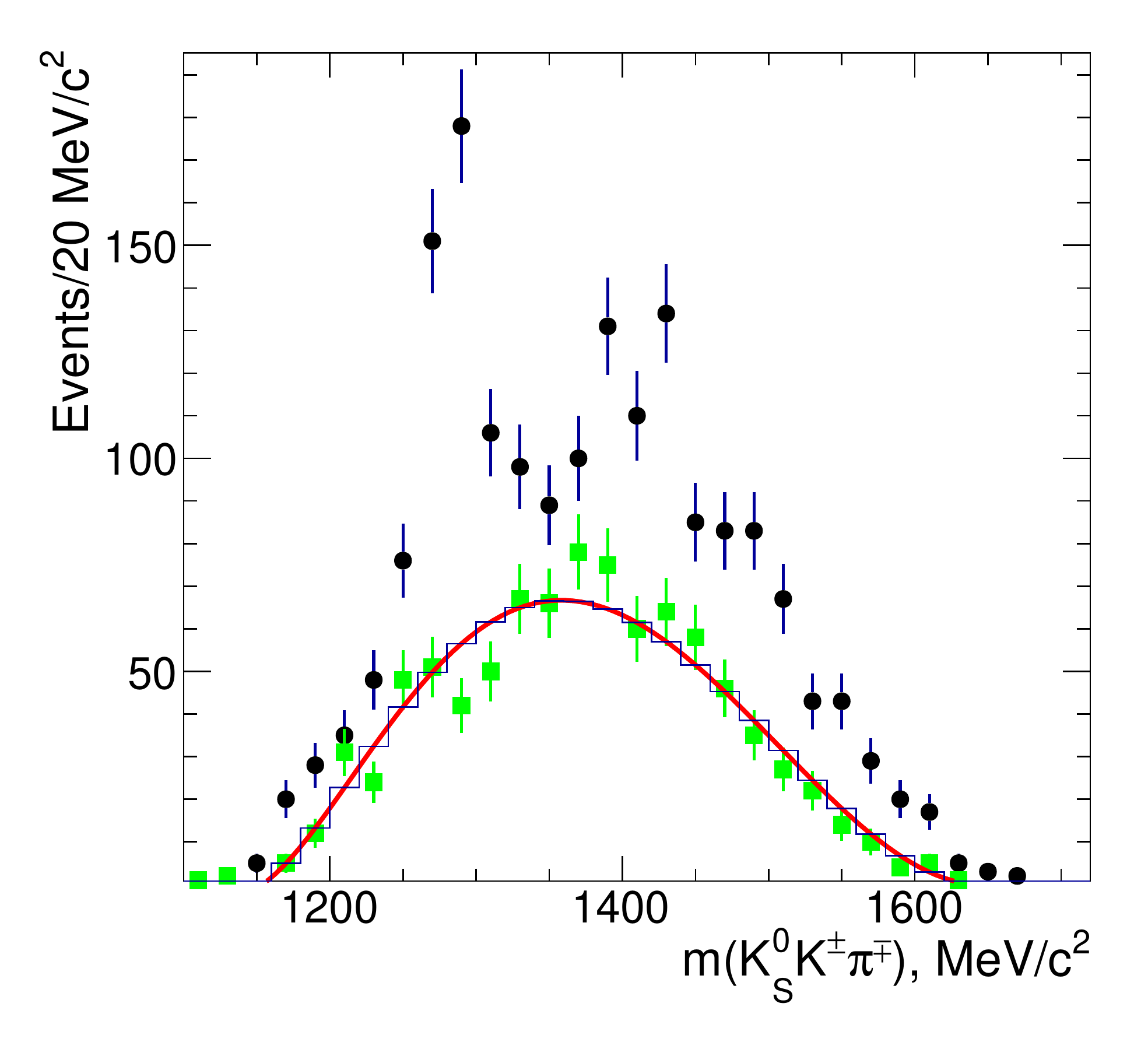}
\put(-35,100){\makebox(0,0)[lb]{\bf(a)}}
\includegraphics[width=0.33\textwidth,height=0.33\textwidth]{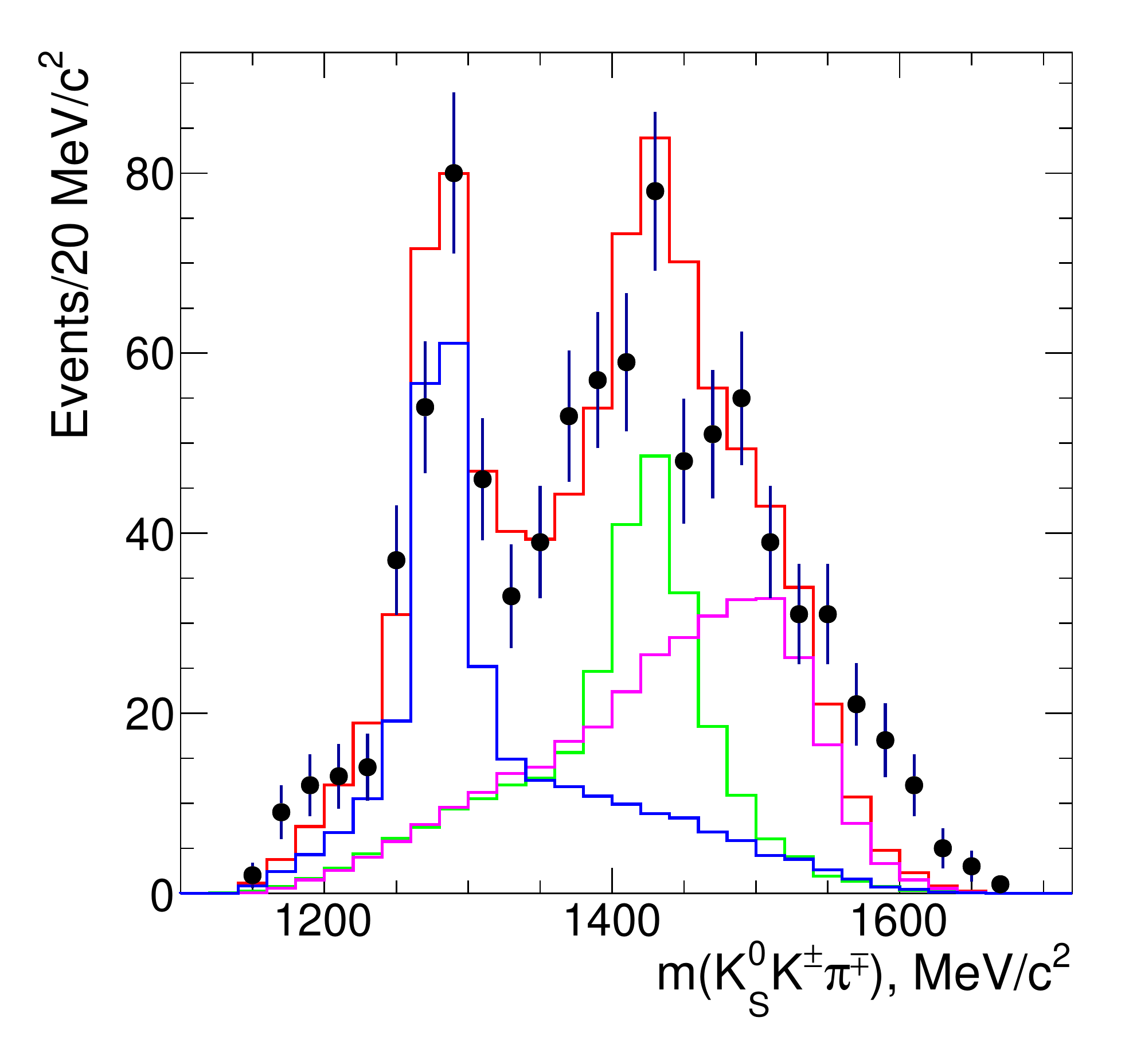}
\put(-35,100){\makebox(0,0)[lb]{\bf(b)}}
\includegraphics[width=0.33\textwidth,height=0.33\textwidth]{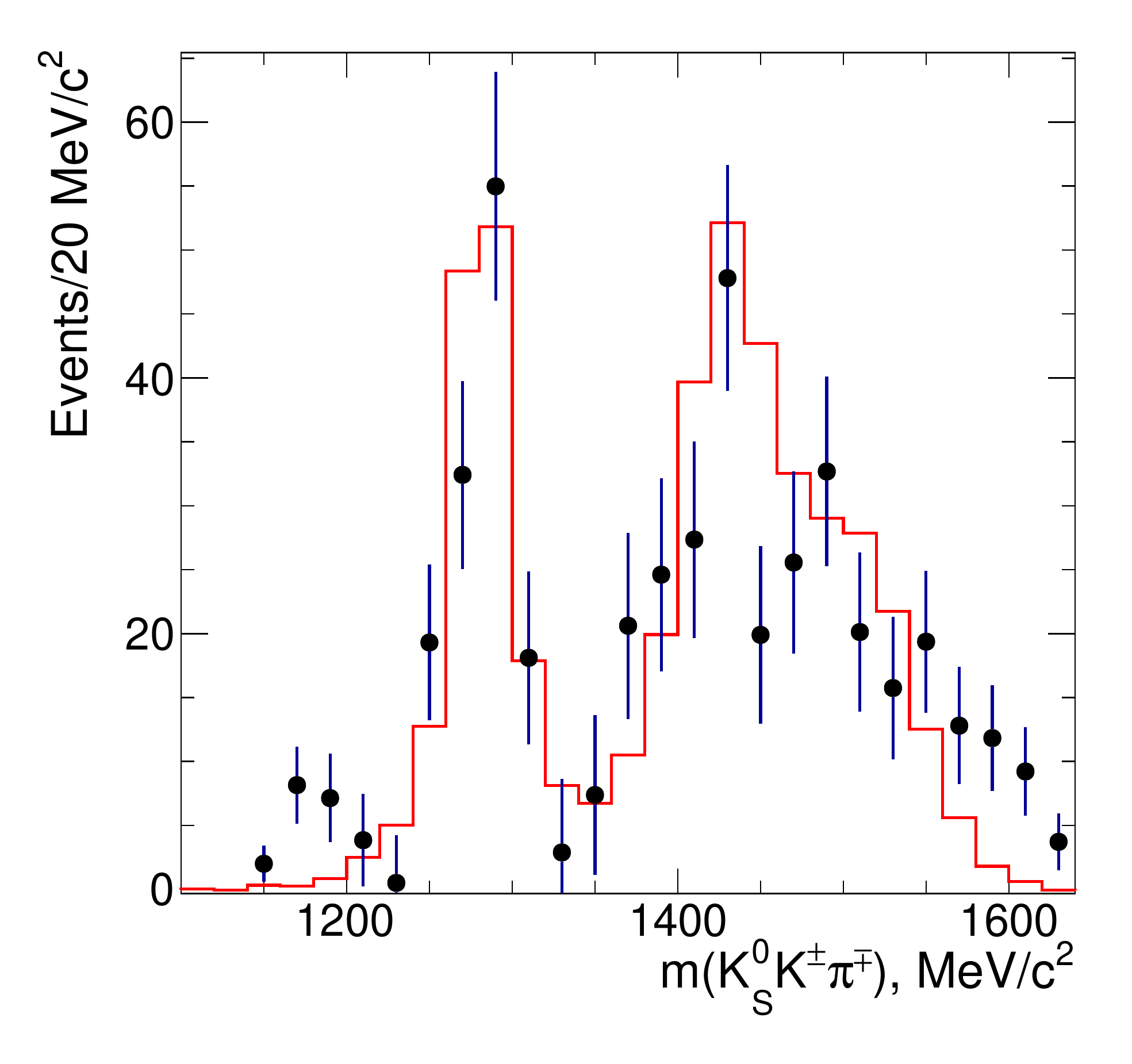}
\put(-35,100){\makebox(0,0)[lb]{\bf(c)}}
\vspace{-0.5cm}
\caption
{(a) The $m(K_S^0K^{\pm}\pi^{\mp})$ invariant mass distribution (dots) 
  (two entries per event, sum for all energy intervals). The curve and
 the histogram show the fit to the combinatorial background, estimated
  from a wrong-sign combination, $m(K_S^0K^{\pm}\pi^{\pm})$
  (squares). (b) The $m(K_S^0K^{\pm}\pi^{\mp})$ invariant mass
  distribution (dots) for $\mathrm{E}_{\mathrm c.m.}=
  \mathrm{2007~MeV}$. Blue, green and magenta histograms are simulated
  contribution from the $f_1(1285)$, $f_1(1420)$ and $f_1(1510)$
  resonances, while red histogram is a sum of them. (c) The same as
  (b) after combinatorial background subtraction. The histogram is the
  sum of the three resonances from the simulation. 
}
\label{masses}
\end{figure}

The events from the side bands of Fig.~\ref{mksk}(a) have no structures,
and are used to subtract the background in the other distributions. 
The background-subtracted invariant mass for
the two kaons and one pion,  $m(K_S^0K^{\pm}\pi^{\mp})$, 
(two entries per event) is shown  in Fig.~\ref{masses}(a) by points: all energy
intervals are summed. We use both six- and five-track events assigning the
missing four-momenta to the lost pion for the latter case.  
 This distribution indicates a presence of a narrow resonance which
 is interpreted as $f_1(1285)$ from the $e^+e^-\to f_1(1285)\rho(770)$
 reaction. A bump at ~1400~MeV/c$^2$ can be interpreted as $f_1(1420)$
 resonance. These resonances were previously studied in proton-proton
 interactions, see for example Ref.~\cite{ppint}, and have relatively
 well determined parameters~\cite{pdg}. The curve and the
  histogram show the fit to the combinatorial background, obtained
  from a wrong-sign combination, $m(K_S^0K^{\pm}\pi^{\pm})$.

  Figure~\ref{masses}(b) shows the same distribution for the events from the
  E$_{\rm c.m.} = 2007~\rm MeV$ energy point in comparison with the MC
  simulation. We simulate production of  the $f_1(1285)$ and
  $f_1(1420)$ resonances separately and  show the expected contribution by blue 
  and green histograms: here and below we perform an approximate weight of the
  simulated events from the different channels to compare with the data for a 
  demonstration. It is seen, that in addition to the $f_1(1285)$ and $f_1(1420)$ 
  we need to add the $e^+e^-\to f_1(1510)\rho$ reaction with approximately the 
  same weight as $f_1(1420)$, pink histogram, to describe the events at higher
  masses. The $f_1(1510)$ resonance has the same decay modes as  $f_1(1420)$, 
  and it has some inconsistency in the parameters~\cite{pdg}, so its mass and 
  width can be varied in a relatively wide range. The solid red histogram shows 
  a sum of these three simulated reactions with approximately equal weights,
  demonstrating that the above reactions describe the observed mass distribution. 

  Figure~\ref{masses}(c) shows the events from Fig.~\ref{masses}(b) when
  the combinatorial background is subtracted by using the wrong-sign mass 
  distribution: the signal from $f_1(1285)$ is seen almost isolated, while 
  the bump at higher masses can be explained as a sum of the $f_1(1510)$ and 
  $f_1(1420)$ resonances, which are wide and highly overlapped. 
  
\begin{figure}[tbh]
\vspace{-0.cm}
\includegraphics[width=1.0\textwidth,height=0.45\textwidth]{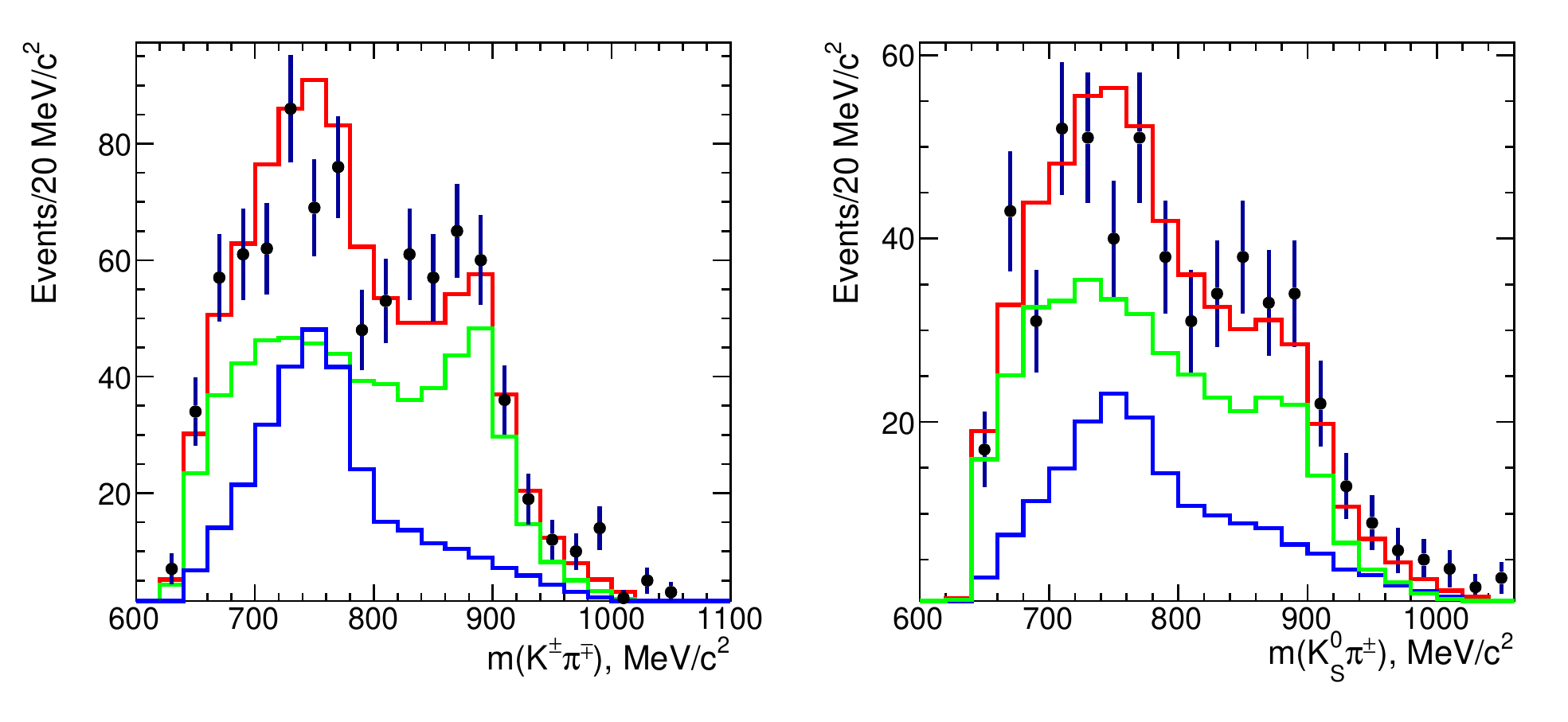}
\put(-235,140){\makebox(0,0)[lb]{\bf(a)}}
\put(-35,140){\makebox(0,0)[lb]{\bf(b)}}
\vspace{-0.5cm}
\caption
{ The experimental $m(K^{\pm}\pi^{\mp})$ (a) and  $m(K_S^0\pi^{\pm})$ (b)
  invariant mass distributions  
  (two entries per event) in comparison with simulation (red histogram)). The blue and green histograms are the MC
  simulation contributions from the $e^+e^-\to f_1(1285)\rho$ and
  $e^+e^-\to f_1(1420,1510)\rho$ reactions, respectively.
}
\label{masses1}
\end{figure}
  
  The existence of the $f_1(1510)$ in addition to $f_1(1420)$ is
  supported by a study of the $K\pi$ invariant mass distributions.
  Figure~\ref{masses1} shows the $m(K^{\pm}\pi^{\mp})$(a) and
  $m(K_S^0\pi^{\pm})$ (b) invariant mass distributions, where signals
  from the $K^*(892)^0$ and $K^*(892)^{\pm}$ are well seen.
  The final state with the $f_1(1285)$ resonance has no $K^*(892)$ signal
  (blue histogram), while the sum of the signals from $f_1(1420)$ and $f_1(1510)$, 
  shown by green histogram, explains the observed experimental distribution. 
  Red histograms present the MC simulation with the sum of the three resonances. 

\begin{figure}[tbh]
\vspace{-0.cm}
\includegraphics[width=1.0\textwidth,height=0.33\textwidth]{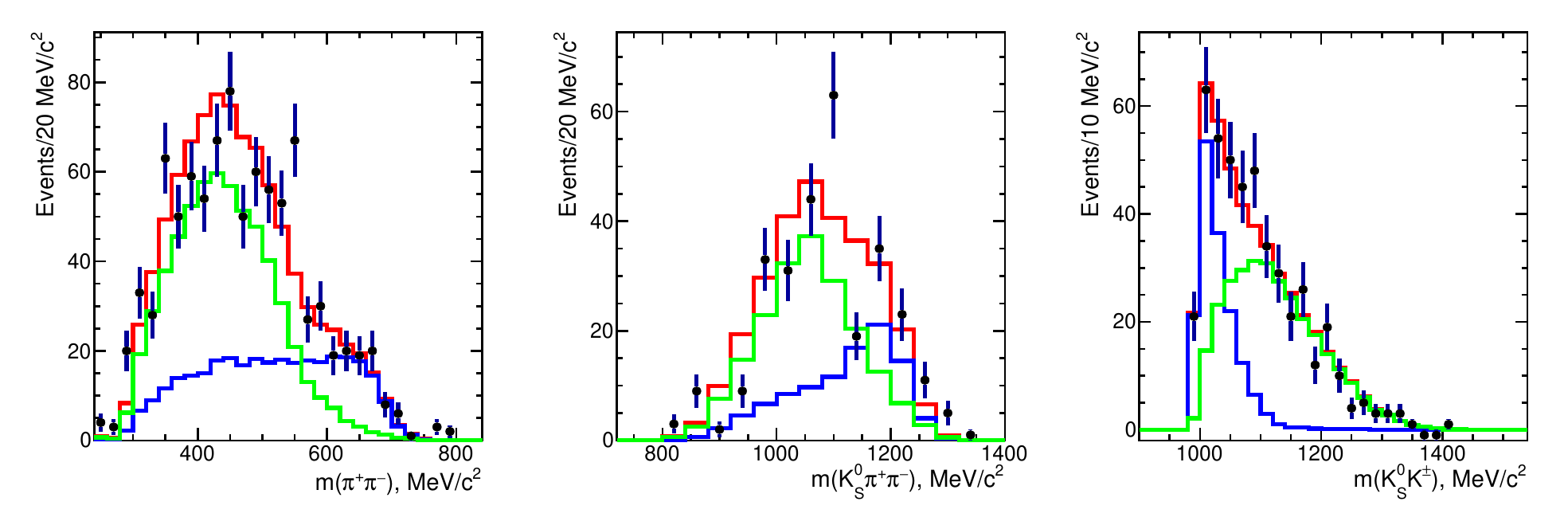}
\put(-295,100){\makebox(0,0)[lb]{\bf(a)}}
\put(-165,100){\makebox(0,0)[lb]{\bf(b)}}
\put(-35,100){\makebox(0,0)[lb]{\bf(c)}}
\vspace{-0.5cm}
\caption
{ The experimental $m(\pi^+\pi^-)$ (a), $m(K_S^0\pi^+\pi^-)$ (b),
  and  $m(K_S^0K^{\pm})$ (c) invariant mass distributions in comparison with the simulation (red histogram).
 The blue and green histograms are the MC
  simulation contributions from the $e^+e^-\to f_1(1285)\rho$ and
  $e^+e^-\to f_1(1420,1510)\rho$ reactions, respectively.
}
\label{masses2}
\end{figure}
  
  Figure~\ref{masses2} shows the background subtracted
  $m(\pi^+\pi^-)$ (a), $m(K_S^0\pi^+\pi^-)$ (b), 
  and  $m(K_S^0K^{\pm})$ (c) invariant mass distributions.  
 The blue and green histograms show the MC simulated contributions
 from the $e^+e^-\to f_1(1285)\rho$ reaction, and  the sum of 
 $e^+e^-\to f_1(1420)\rho$ and  $e^+e^-\to f_1(1510)\rho$ reactions,
 respectively. An enhancement at the large values of $m(\pi^+\pi^-)$
 is due to  a presence of the $\rho(770)$  from the $e^+e^-\to f_1(1285)\rho$ 
 reaction, while in the case of $f_1(1420)$ and $f_1(1510)$ the signal 
 from $\rho(770)$ is suppressed kinematically at our energies due to a 
 large energy fraction taken by the $KK^*$ pair.
 Note, that the $f_1(1285)$ resonance decays to $K\bar K\pi$ via
  the $a_0(980)\pi$ intermediate state, while the $f_1(1510)$ and
  $f_1(1420)$ resonances  both predominantly decay to the $KK^*(892)$ mode.
 A dominance of the $f_1(1285)\to a_0(980)\pi\to K\bar K\pi$
 intermediate reaction is demonstrated in Fig.~\ref{masses2}(c), where
 the $K_S^0 K^{\pm}$ invariant masses are concentrated at the threshold and
 well described by the simulation.

 From the above study we conclude that the $e^+e^-\to K_S^0
 K^{\pm}\pi^{\mp}\pi^+\pi^-$ process dominates by the $e^+e^-\to
 f_1\rho(770)$ reaction with a combination of the $f_1(1285)$,
 $f_1(1420)$, and $f_1(1510)$ resonances. A small contribution 
 from the other non-resonant processes is not excluded. 

 The above reactions have the same final state and an interference of
  them could change the observed mass distributions. The influence of the
  interference to the $m(K_S^0K^{\pm}\pi^{\mp})$ distribution is
  studied with the MC simulation.  Because of the relatively narrow width and the difference in the
  intermediate state, the $f_1(1285)$ signal has practically no
  influence from the interference with  the $f_1(1420)$ and $f_1(1510)$
  resonances, while the latter two are highly overlapped and the mass shape
  depends on the relative phase between the amplitudes.

\begin{center}
\begin{figure}[tbh]
\vspace{-0.2cm}
\includegraphics[width=1.0\textwidth]{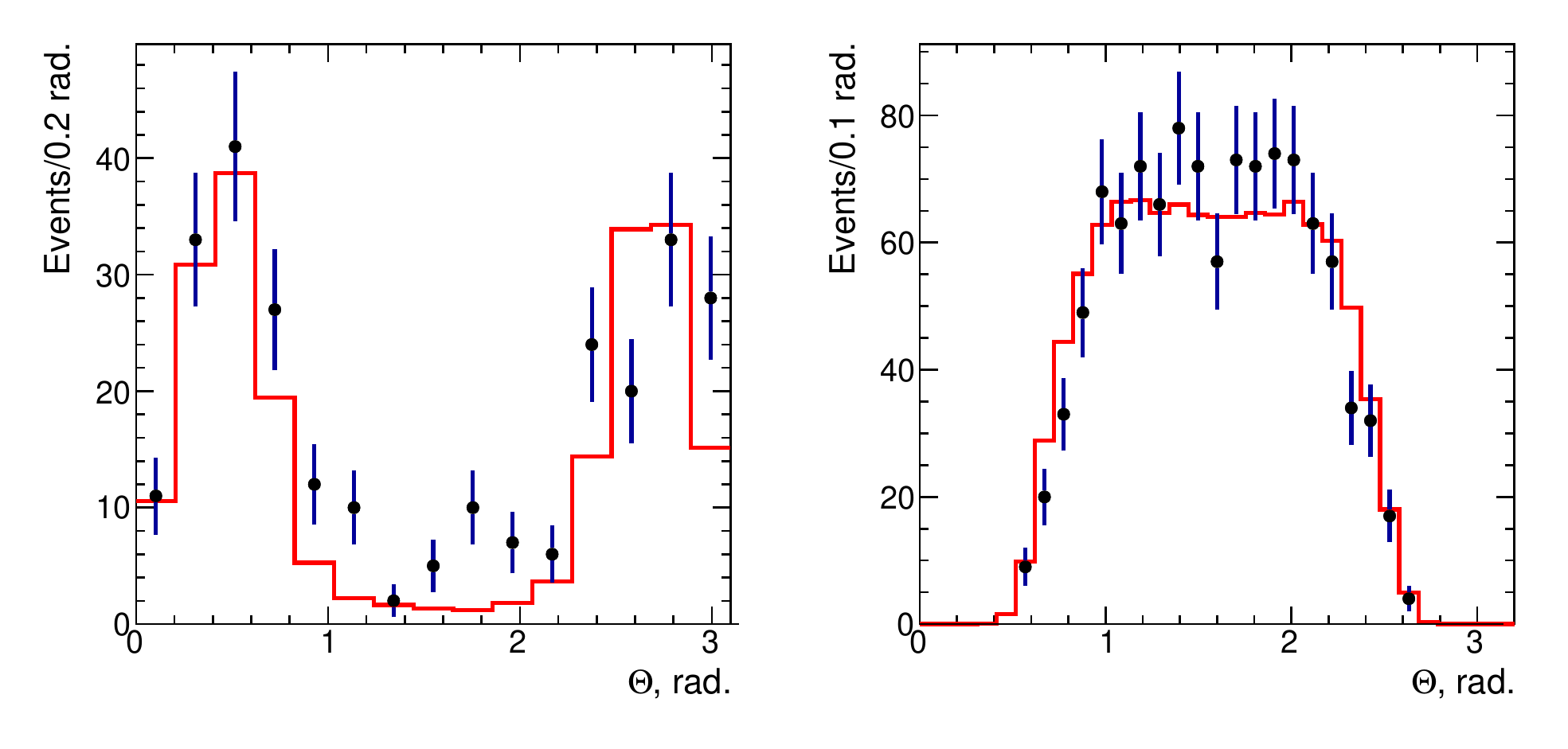}
\put(-245,153){\makebox(0,0)[lb]{\bf(a)}}
\put(-45,153){\makebox(0,0)[lb]{\bf(b)}}
\vspace{-0.8cm}
\caption
{(a) The background-subtracted experimental (dots) polar angle, $\Theta$,
  distribution  
in comparison with the simulated distribution (histograms,
$f_1\rho(770)$ model) for the missing 
pion (a) and all detected pions (b).
}
\label{5/6tracks}
\end{figure}
\end{center}
\begin{center}
\begin{figure}[tbh]
\vspace{-0.cm}
\includegraphics[width=1.\textwidth]{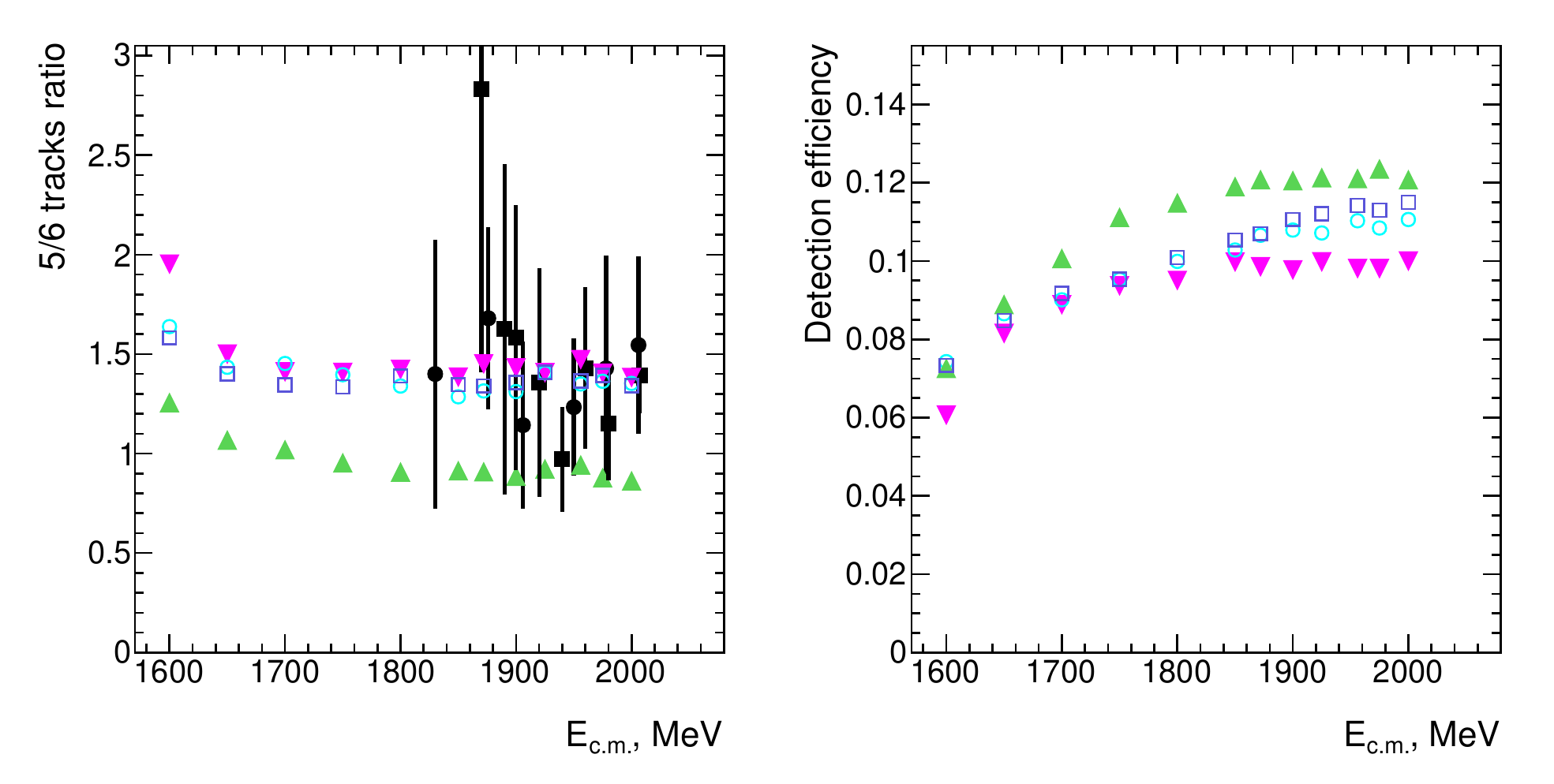}
\put(-245,163){\makebox(0,0)[lb]{\bf(a)}}
\put(-45,163){\makebox(0,0)[lb]{\bf(b)}}
\vspace{-0.5cm}
\caption
{ (a) The ratio of the number of five- to six-track events for the data (dots) 
and simulation for the different intermediate states: PS
model (triangles), $f_1(1285)\rho$ 
(up-down triangles), $f_1(1420)\rho$ (open circles), and
$f_1(1510)\rho$  (open squares) intermediate states.
(b) The detection efficiency obtained from the MC simulation for the 
$e^+ e^-\to K_S^0 K^{\pm}\pi^{\mp}\pi^+\pi^-$ reaction for the different
intermediate states (symbols legend is the same as for (a)).
}
\label{efficiency}
\end{figure}
\end{center}
\section{Detection efficiency}
\label{sec:efficiency}
\hspace*{\parindent}

Since DC acceptance in our experiment is about 70\%, the detection 
efficiency depends on the particle angular distribitions determined by 
the hadron production dynamics. In addition, we have to take into account 
some minor track reconstruction inefficiency. 

To obtain the detection efficiency value, we simulate $K_S^0 K^{\pm}\pi^{\mp}\pi^+\pi^-$ 
production in the primary generators, 50000 events for each c.m. energy 
interval for each model, 
trace simulated particles through the 
\mbox{CMD-3} detector using the GEANT4~\cite{geant4} package, and reconstruct them
with the same software as experimental data. 
We calculate the detection efficiency from the MC-simulated events
as a ratio of events after the selections described in 
Secs.~\ref{select}, \ref{dynamics} to the total number of generated events.

Our selection of the six- and the five-track signal events allows us to estimate a 
difference in the tracking efficiency in the data and simulation, and perform a test of the model
used for the efficiency calculation.
Figure~\ref{5/6tracks} shows by dots the background-subtracted polar angle 
for a missing pion (a) and for all detected charged tracks (b). Solid histograms represent
the simulated distribution for the $f_1\rho$ intermediate state,
normalized to the number of experimental events in Fig.~\ref{5/6tracks}(b).
The events inside the DC acceptance, seen in the
1.0--2.1 radians region of Fig.~\ref{5/6tracks}(a), are used to estimate the difference
in the track resonstruction efficiency for data and simulation (see below).

The calculated ratio of the number of five- to six-track events 
at each c.m. energy interval is shown in Fig.~\ref{efficiency}(a) by points for the data.
The values of the ratio for the PS model (triangles), and
for the $f_1(1285,1420,1510)\rho$ intermediate states 
(up-down triangles, open squares, and open circles, respectively) are
 shown in Fig.~\ref{efficiency}(a) for a comparison. The PS model is not
compatible with our data.

We calculate the detection efficiency for a sum of the 
events with the six and five detected tracks.
Figure~\ref{efficiency}(b) shows the detection efficiency obtained for 
the $e^+ e^-\to K_S^0 K^{\pm}\pi^{\mp}\pi^+\pi^-$ reaction for different intermediate 
states: markers are the same as for Fig.~\ref{efficiency}(a).

\begin{center}
\begin{figure}[tbh]
\vspace{-0.2cm}
\includegraphics[width=0.5\textwidth]{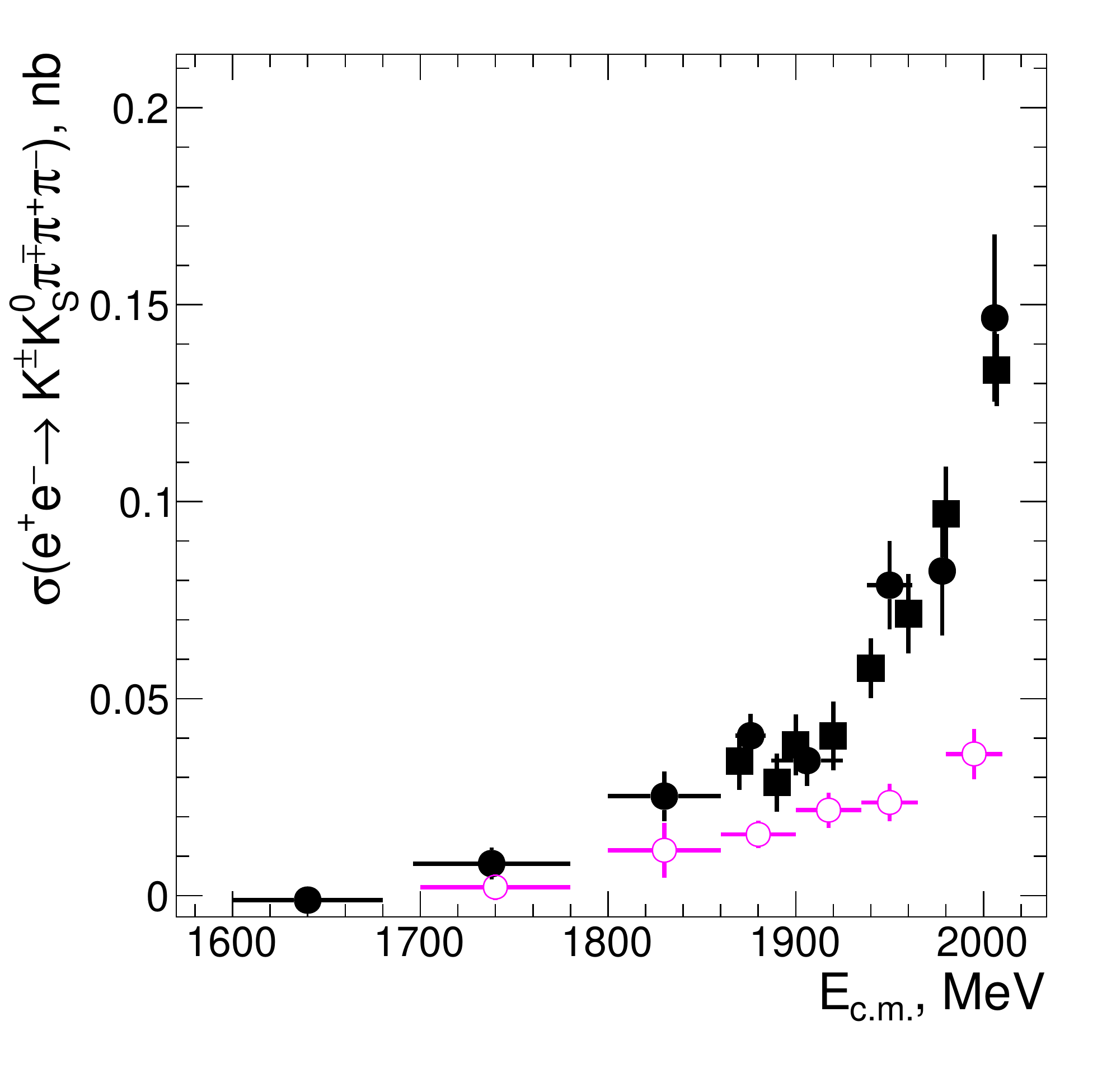}
\put(-145,153){\makebox(0,0)[lb]{\bf(a)}}
\includegraphics[width=0.5\textwidth]{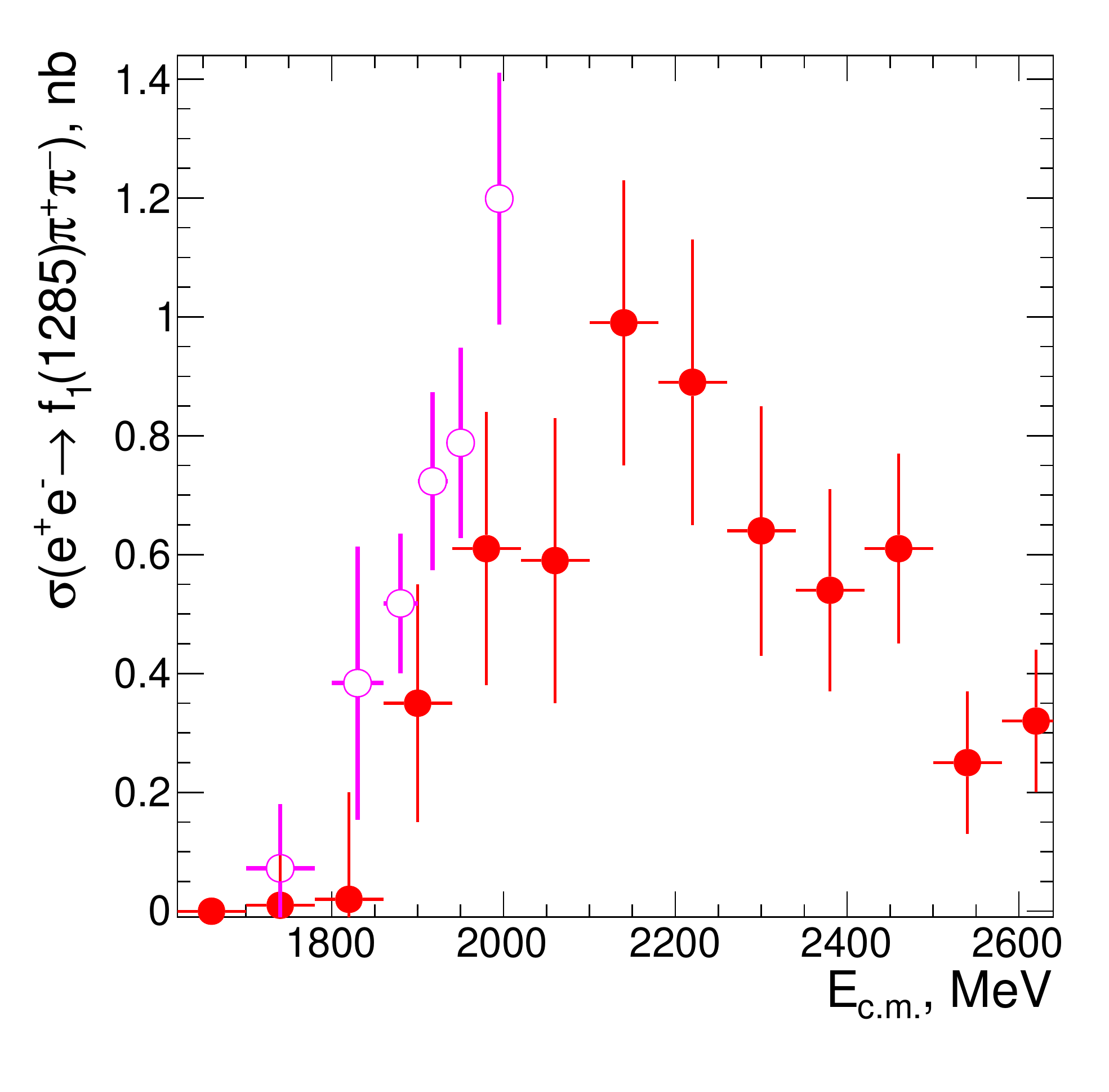}
\put(-45,153){\makebox(0,0)[lb]{\bf(b)}}
\vspace{-0.5cm}
\caption{
(a) The $e^+e^-\to K_S^0 K^{\pm}\pi^{\mp}\pi^+\pi^-$ cross section measured with the \mbox{CMD-3}
detector at \mbox{VEPP-2000} (filled circles for combined scans, filled
squares for individual energy points). 
The cross section for the $e^+e^-\to f_1(1285)\pi^+\pi^-\to K_S^0 K^{\pm}\pi^{\mp}\pi^+\pi^-$ reaction 
is shown by open circles.
(b) The $e^+e^-\to f_1(1285)\pi^+\pi^-$ cross section with the correction
for the missing decay modes (open circles) in comparison with BaBar
data~\cite{isr5pi} measured in the $f_1(1285)\to\eta\pi^+\pi^-$ mode.
}
\label{cross}
\end{figure}
\end{center}
\section{Cross section calculation}    
\hspace*{\parindent}
\label{xs_tot}
In each energy interval the cross section is calculated as 
$$
\sigma = \frac{N_{6tr}+N_{5tr}}{L\cdot\epsilon\cdot(1+\delta)},
$$ 
where $N_{6tr}$, $N_{5tr}$ are the background-subtracted numbers of 
the signal events with six and five tracks, $L$ is the integrated 
luminosity for this energy interval, $\epsilon$ is the detection efficiency, 
and $(1+\delta)$ is a radiative correction calculated according to 
Ref.~\cite{kur_fad,actis}. To calculate the radiative correction, 
we use a procedure with the iterations of the observed cross section 
in the radiative integral, and obtain $(1+\delta) = 0.85$ with a weak 
energy dependence.  

As shown above the observed events are the mixture of the different
intermediate states with slightly different detection efficiency. For
the inclusive $e^+e^-\to K_S^0 K^{\pm}\pi^{\mp}\pi^+\pi^-$  cross
section we use an averaged value of the efficiency for the three observed
modes. We assign $\pm$5\% systematic uncertainty to this procedure,
which is close to the difference between calculated efficiencies for
the observed reactions. 
We also apply
some corrections due to the difference in the data and MC simulation for
the tracking efficiency (see Sec.~\ref{systematic}).
The cross section is shown in Fig.~\ref{cross}(a) by filled circles for earlier runs, combined
into the eight energy intervals, and filled squares for the latest scans. There are no other measurements.
We also calculate the cross section by using only events with the six detected tracks: 
a less than 5\% difference within statistical fluctuation is observed.

Energy interval, integrated luminosity, the number of six- and five-track 
events, and the obtained cross section for each energy interval 
are listed in Table~\ref{xs_all}. 

\begin{table}[tbh]
\caption{Energy interval, integrated luminosity, number of signal 6-track events,  number of signal 5-track events, 
 and the obtained cross section for the $e^+e^-\to K_S^0 K^{\pm}\pi^{\mp}\pi^+\pi^-$ reaction.
Only statistical uncertainties are shown. 
}
\label{xs_all}
\vspace{-0.cm}
\begin{center}
\renewcommand{\arraystretch}{0.85}
\begin{tabular}{cccccccc}
\hline
  {E$_{\rm c.m.}$, MeV}
 & {$L$, nb$^{-1}$} 
&{$N_{6tr}$}
&{$N_{5tr}$}
&{$\sigma_{K_S^0 K^{\pm}\pi^{\mp}\pi^+\pi^-}$, nb}\\

  \hline
2000--2007 & 4259 & 22.0 $\pm$ 4.7 & 34.0 $\pm$ 6.6 & 0.147 $\pm$ 0.021\\ 
1975--1980 & 4640 & 14.0 $\pm$ 4.0 & 20.0 $\pm$ 5.5 & 0.082 $\pm$ 0.016\\ 
1940--1962 & 9653 & 30.0 $\pm$ 5.7 & 37.0 $\pm$ 7.7 & 0.079 $\pm$ 0.011\\ 
1890--1925 & 15158 & 21.0 $\pm$ 4.8 & 24.0 $\pm$ 6.9 & 0.034 $\pm$ 0.006\\ 
1870--1884 & 19333 & 25.0 $\pm$ 5.0 & 42.0 $\pm$ 7.7 & 0.041 $\pm$ 0.006\\ 
1800--1860 & 11428 & 10.0 $\pm$ 3.2 & 14.0 $\pm$ 5.1 & 0.025 $\pm$ 0.006\\ 
1700--1780 & 12783 & 1.0 $\pm$ 1.7 & 7.0 $\pm$ 3.6 & 0.008 $\pm$ 0.004\\ 
1600--1680 & 13193 & 0.0 $\pm$ 0.0 & --1.0 $\pm$ 1.7 & --0.001 $\pm$ 0.002\\
2007 & 21643 & 107.0 $\pm$ 10.5 & 152.0 $\pm$ 14.2 & 0.133 $\pm$ 0.009\\ 
1980 & 10093 & 39.0 $\pm$ 6.6 & 48.0 $\pm$ 8.6 & 0.097 $\pm$ 0.012\\ 
1960 & 11065 & 27.0 $\pm$ 5.9 & 43.0 $\pm$ 7.9 & 0.072 $\pm$ 0.010\\ 
1940 & 14416 & 35.0 $\pm$ 5.9 & 38.0 $\pm$ 7.6 & 0.058 $\pm$ 0.008\\ 
1920 & 9904 & 14.0 $\pm$ 3.7 & 21.0 $\pm$ 6.6 & 0.041 $\pm$ 0.009\\ 
1900 & 9675 & 13.0 $\pm$ 3.6 & 19.0 $\pm$ 5.4 & 0.038 $\pm$ 0.008\\ 
1890 & 8912 & 8.0 $\pm$ 2.8 & 14.0 $\pm$ 4.9 & 0.029 $\pm$ 0.007\\ 
1870 & 9286 & 6.0 $\pm$ 2.4 & 21.0 $\pm$ 5.2 & 0.034 $\pm$ 0.007\\ 

\hline
\end{tabular}
\end{center}
\end{table}

\section{Cross section of the $e^+e^-\to f_1(1285)\pi^+\pi^-$}    
\hspace*{\parindent}
\label{xs_f1285}

The distributions shown in Fig.~\ref{masses}(c) can be used to  extract the number of
events associated with the $e^+e^-\to f_1(1285)\pi^+\pi^-$ reaction.
After the combinatorial background subtraction, the $f_1(1285)$ signal
is well seen  and has low remaining non-resonant background.
Simulation shows that a possible interference between $f_1(1285)$ and
$f_1(1420)$--$f_1(1510)$ resonances is negligible because of the
difference in the width and in the decay modes.   
The numbers of events are
extracted in the $\pm 30$~MeV/c$^2$ mass interval around $f_1(1285)$
peak of Fig.~\ref{masses}(c) with a background subtraction from  side
bands. The numbers of events are small therefore  we combine the
results into the six energy intervals. 
The results for the number of events in the energy intervals and the
integrated luminosity are shown in Table~\ref{xs_f1}.  
The contribution of the  $e^+e^-\to f_1(1285)\pi^+\pi^-$ reaction to 
the inclusive  $e^+e^-\to K_S^0 K^{\pm}\pi^{\mp}\pi^+\pi^-$ cross
section is shown in  Fig.~\ref{cross}(a) and listed in
Table~\ref{xs_f1}. 

 We correct the number of events for the missing
decay modes of the $f_1(1285)$ resonance, containing $K_L^0$ and $\pi^0$ (factor of 3), and using the
branching fraction of $f_1(1285)\to K\bar K\pi$ from Ref.~\cite{pdg}
we obtain the cross section for the  $e^+e^-\to f_1(1285)\pi^+\pi^-$
reaction. The obtained cross section for the $e^+e^-\to
f_1(1285)\pi^+\pi^-$
 reaction is shown in Fig.~\ref{cross}(b) by open circles and listed in Table~\ref{xs_f1}.

This cross section can be compared with the only available
measurement by BaBar~\cite{isr5pi}, shown in Fig.~\ref{cross}(b), in which the  $f_1(1285)$
resonance was observed in the $\eta\pi^+\pi^-$ decay mode. Our
measurement demonstrates a faster rise of the cross section from the threshold.

Because of the relatively large widths, large uncertainty in the
parameters, and unknown influence of the interference we cannot extract
separately the events for the $f_1(1420)$ and $f_1(1510)$ resonances
from our data. 

\begin{table}[tbh]
\caption{Energy interval, integrated luminosity, number of signal
  events, the obtained cross section for the 
$e^+e^-\to f_1(1285)\pi^+\pi^-\to K_S^0 K^{\pm}\pi^{\mp}\pi^+\pi^-$
reaction, and for the $e^+e^-\to f_1(1285)\pi^+\pi^-$ process.
Only statistical uncertainties are shown. 
}
\label{xs_f1}
\vspace{-0.cm}
\begin{center}
\renewcommand{\arraystretch}{0.85}
\begin{tabular}{cccccccc}
\hline
  {E$_{\rm c.m.}$, MeV}
  & {$L$, nb$^{-1}$} 
&{$N_{f_1}$}
&{$\sigma_{K_S^0 K^{\pm}\pi^{\mp}\pi^+\pi^-}$, nb}
&{$\sigma_{f_1(1285)\pi^+\pi^-}$, nb}\\

  \hline
1700--1780 & 12783 & 2.0 $\pm$ 40.0 & 0.002 $\pm$ 0.003 & 0.07 $\pm$ 0.11\\ 
1800--1860 & 11428 & 10.0 $\pm$ 30.0 & 0.012 $\pm$ 0.007 & 0.38 $\pm$ 0.23\\ 
1860--1900 & 36903 & 44.0 $\pm$ 20.0 & 0.016 $\pm$ 0.004 & 0.52 $\pm$ 0.12\\ 
1900--1935 & 34738 & 58.0 $\pm$ 17.5 & 0.022 $\pm$ 0.004 & 0.72 $\pm$ 0.15\\ 
1935--1965 & 35134 & 64.0 $\pm$ 15.0 & 0.024 $\pm$ 0.005 & 0.79 $\pm$ 0.16\\ 
1980--2010 & 40675 & 113.0 $\pm$ 15.0 & 0.036 $\pm$ 0.006 & 1.20 $\pm$ 0.21\\   
\hline
\end{tabular}
\end{center}
\end{table}

\section{Systematic uncertainties}
\hspace*{\parindent}
\label{systematic}
The following sources of systematic uncertainties and corrections are
considered for the cross section measurement.

\begin{itemize}

\item{The major uncertainty in the event number comes from the
    separation of the kaons and pions using the dE/dX values. As shown in
    Fig.~\ref{mksk}(c) signals from the kaons and pions are highly
    overlapped, and the applied boundary rejects about 50\% of the kaons
    with momentum above 300 MeV/c, corresponding to  about 20\% of losses
    in the total number of the events. 
    By changing the boundary we either loosing the signal
    events or rapidly increase the pion leakage to the kaons.
    We vary the boundary  and only half of these losses is properly
    described by the simulation.
    It corresponds to  about $\pm$10\% systematic
    uncertainty in the result. 
}
  
\item{The tracking efficiency was studied in detail in our previous 
    papers~\cite{cmd6pi,cmd4pi}. A similar estimate is made using
    angular distribution in Figs.~\ref{5/6tracks}(a,b) where we observe a
    difference for the data and MC simulation in the number of missing
    tracks inside the DC acceptance. 
    The correction for the track reconstruction 
efficiency compared to the MC simulation is estimated as about (2.0$\pm$1.0)\% per track.
Since we add events with one missing pion track,  
the MC-simulated detection efficiency is corrected by (--7$\pm$5)\%: 
the uncertainty is taken as the corresponding systematic uncertainty. 
}
\item {
The model dependence of the acceptance is determined by a comparing 
efficiencies calculated for the different production 
dynamics. Since we assume a mixture of the $f_1(1285,1420,1510)\rho$ intermediate
states with the approximately equal contributions, we average the efficiences shown in Fig.~\ref{efficiency}, and
assign a 5\% uncertainty to the calculation.
}  
\item{
Since only one charged track is 
sufficient for a trigger (98--99\% single track efficiency), and using
a cross check with the independent neutral trigger, we conclude 
that the trigger inefficiency for the multitrack events gives a negligible contribution to the systematic 
uncertainty. 
}
\item{
The systematic uncertainty due to the selection criteria is studied by 
varying the requirements described above and doesn't exceed 5\%. 
}
\item{
The uncertainty on the the integrated luminosity 
comes from the selection criteria of the Bhabha events, the radiative
corrections, the detector calibrations, and does not exceed 
1.5\%~\cite{lum}.
}
\item{
The radiative correction uncertainty is estimated as about 
2\%, mainly due to the uncertainty on the maximum allowed energy of the 
emitted photon, as well as from the uncertainty on the cross section.
}
\end{itemize}

The above systematic uncertainties summed in quadrature give an overall
systematic uncertainty of about 15\%.

\section*{ \boldmath Conclusion}
\hspace*{\parindent}
The total cross section of the process $e^+e^-\to K_S^0 K^{\pm}\pi^{\mp}\pi^+\pi^-$ 
has been measured using 185.4 pb$^{-1}$ of integrated 
luminosity collected by the \mbox{CMD-3} detector at the \mbox{VEPP-2000} $e^+e^-$ collider
in the 1.6--2.0~GeV c.m. energy range. The systematic uncertainty is 
about 15\%. At the present statistical accuracy we do not observe any
influence of the $N\bar N$ threshold to the cross section. 
From our study  we can conclude that the observed final state can be 
described by the 
$e^+e^-\to f_1\pi^+\pi^-$ reaction with the contribution from the $f_1(1285)$,
$f_1(1420)$, and $f_1(1510)$ resonances. We extracted the number of events associated
with the $f_1(1285)$ resonance, and
the measured cross section for the $e^+e^-\to f_1(1285)\pi^+\pi^-$ reaction  
is compatible with the only available measurement by BaBar.

\subsection*{Acknowledgments}
\hspace*{\parindent}
We thank the \mbox{VEPP-2000} team for the excellent machine operation. 
The work is partially supported by the Russian 
Foundation for Basic Research grant 20-02-00496.

\end{document}